 \documentclass[preprint]{aastex}

\shorttitle{Deep CCD Photometry of NGC 6819}
\shortauthors{Kalirai, J. S. \it et al. \normalfont}

\begin{document}


\title{The CFHT Open Star Cluster Survey II -- Deep CCD
Photometry of the Old Open Star Cluster NGC 6819}


\author{Jasonjot Singh Kalirai\altaffilmark{1}}
\affil{Physics \& Astronomy Department, 6224 Agricultural Road,
University of British Columbia, Vancouver, BC V6T-1Z1}
\email{jkalirai@physics.ubc.ca}

%\author{Harvey B. Richer, Gregory G. Fahlman,
%Francesca D'Antona}

\author{Harvey B. Richer\altaffilmark{1}}
%\affil{University of British Columbia}
\author{Gregory G. Fahlman \altaffilmark{1,2}}
%\affil{Canada-France-Hawaii Telescope}
\author{Jean-Charles Cuillandre \altaffilmark{2}}
%\affil{Canada-France-Hawaii Telescope}
\author{Paolo Ventura \altaffilmark{3}}
%\affil{Osservatorio Astronomico di Roma}
\author{Francesca D'Antona \altaffilmark{3}}
%\affil{Osservatorio Astronomico di Roma}
\author{Emmanuel Bertin \altaffilmark{4}}
%\affil{Institut D'Astrophysique De Paris}
\author{Gianni Marconi \altaffilmark{5}}
%\affil{ESO}

\and

\author{Patrick R. Durrell \altaffilmark{6}}
%\affil{Penn. State University}


\altaffiltext{1}{University of British Columbia}
\altaffiltext{2}{Canada-France-Hawaii Telescope Corporation}
\altaffiltext{3}{Osservatorio Astronomico di Roma}
\altaffiltext{4}{Institut D'Astrophysique De Paris}
\altaffiltext{5}{European Southern Observatory}
\altaffiltext{6}{Penn. State University}


\begin{abstract}
We present analysis of deep CCD photometry for the very rich, old
open star cluster NGC 6819.  The science goals are to catalogue
the white dwarfs in the cluster and measure the cluster luminosity
and mass functions. These CFH12K data results represent the first
of nineteen open star clusters which were imaged as a part of the
CFHT Open Star Cluster Survey. We find a tight, very rich,
main-sequence and turn-off consisting of over 2900 cluster stars
in the V,  B$\rm-V$ color-magnitude diagram (CMD). Main-sequence
fitting of the un-evolved cluster stars with the Hyades star
cluster yields a distance modulus of (m$\rm-$M)$_{V}$ = 12.30
$\pm$ 0.12, for a reddening of E(B$\rm-$V) = 0.10. These values
are consistent with a newly calculated theoretical stellar
isochrone of age 2.5 Gyrs, which we take to be the age of the
cluster. Both the depth gained in the photometry and the increased
projected area of the CFH12K Mosaic CCD allow for detailed star
counts in concentric annuli out to large angular radii.  These
indicate a much larger cluster extent (R = 9$\farcm$5 $\pm$
1$\farcm$0), by a factor of $\sim$2 over some previous estimates.
Incompleteness tests confirm a slightly negatively sloped
luminosity function extending to faint (V $\sim$ 23) magnitudes
which is indicative of a dynamically evolved cluster. Further
luminosity function and mass segregation tests indicate that low
mass objects (M $\leq$ 0.65M$_\odot$) predominate in the outer
regions of the cluster, 3$\farcm$5 $\leq$ R $\leq$ 9$\farcm$5. The
estimation of the number of white dwarfs in NGC 6819, based on
stellar evolution models, white dwarf cooling timescales, and
conservation of star number arguments applied to the red giant
stars of the cluster are in good agreement with the observed
number. For those white dwarf candidates which pass both a
statistical subtraction that removes background galaxies and field
stars, and a high star/galaxy confidence by using image
classification, we show comparisons to white dwarf isochrones and
cooling models which suggest the need for spectroscopy to confirm
the white dwarf nature of the brighter objects. This is entirely
feasible for all objects, before a statistical subtraction cut,
with the current generation of 8 meter class telescopes and
multi-object spectrometers.

--------------------------------------

\end{abstract}




\section{Introduction} \label{intro}

    The primary goal of the CFHT Open Star Cluster Survey is to
catalogue a large number of white dwarf stars and provide
observational constraints to the theoretical models of
intermediate and young stellar clusters and their inhabitants (see
\S 1 of Kalirai et al. 2001a, hereafter Paper I). These models,
such as those for the initial-final mass relationship of the
progenitor white dwarf star, or the upper mass limit to white
dwarf production, have never been tested with the detail that is
now possible. Studies involving white dwarf stars in open clusters
have been limited for four major reasons: (1) the majority of
these clusters are not old enough to have produced a sizeable
white dwarf population, (2) most are not rich enough to contain
numerous white dwarfs, (3) these clusters lie in the plane of the
Galaxy so that foreground and background contamination is high,
and (4) the photometric depth of most previous studies is not deep
enough to clearly see white dwarfs.  The first two factors result
in very few white dwarfs, often quite scattered somewhere in the
faint, blue end of the CMD, and the third results in a large
amount of contamination from both field stars and background
galaxies, which also appear as faint, blue objects. The fourth
factor has always been a deterrent to the serious study of cluster
white dwarfs. White dwarf cooling models (Wood 1994; Bergeron
1995), indicate that a significant number of white dwarfs are not
expected brighter than an absolute magnitude of M$_{V} \sim$ 10.
At the opposite end, it is known that very old, cool white dwarfs
may reach M$_{V} \sim$ 17.5. Even the bright end of the white
dwarf cooling sequence is often too faint for the limiting
magnitude that could possibly be reached by many telescopes, even
for moderately close clusters. The new large field detectors and
imagers on 4 m class telescopes are ideal instruments to study
these faint magnitudes in open star clusters. This was
demonstrated in the white dwarf analysis of M67 \citep{richer2}.
The current data set for NGC 6819 is ideal for white dwarf studies
because the cluster is both old and rich, and the photometry
extends to faint, V $\sim$ 25 magnitude, stars.

    NGC 6819 was identified as an old stellar system almost 30 years
ago (Lindoff 1972; Auner 1974).  These early studies used
photographic plates or photoelectric detectors and concentrated on
bright stars. Cluster ages were determined by calibrating relative
positions of turn-off and red giant branch stars on the CMD and by
comparing the bright stars in the cluster to those of the evolved
system M67. These methods produced a large range of ages for the
cluster: 2 Gyrs \citep{lindoff} -- 4 Gyrs \citep{kaluzny}. More
recently, a detailed isochrone fit to the cluster photometry has
been carried out by Rosvick and VandenBerg (1998), and a new age
estimate of 2.4 Gyrs has been determined using models with a
significant amount of convective core overshooting (the age
dependent on the amount of overshooting). The limiting magnitude
of their CCD photometry (V $\sim$ 18.5), the deepest for this
cluster at the time, is too bright to include the low mass
main-sequence stars and far too bright to detect any cluster white
dwarfs.

    Isochrone fitting has been commonly used to constrain the age
measurement for star clusters.  The difficulty is often in
determining key parameters for the cluster such as the reddening,
distance modulus, and metallicity.  The uncertainty in these
values can lead to an ambiguous isochrone fit.  For NGC 6819, the
reddening value causes the largest concern as E(B$\rm-$V) has
ranged in the literature from 0.30 (Lindoff 1972; Auner 1974), to
E(B$\rm-$V) = 0.12 \citep{burkhead}.  The earlier high estimates
have been acknowledged as being derived from poor U filter
photographic photometry.  More recently, a reddening value of 0.15
\citep{canterna} from photoelectric observations was found to
agree with that obtained by Burkhead (1971).  The latest value is
given as E(B$\rm-$V) = 0.16 for the reddening, and
(m$\rm-$M)$_{V}$ = 12.35 for the distance modulus \citep{rosvick}.
This reddening value has been determined by first comparing the
red giant clump of NGC 6819 with that of M67, and then adjusting
the value slightly based on the theoretical isochrone fits.
Additionally, a recent spectroscopic study of the red clump stars
in NGC 6819 has yielded a reddening value of E(B$\rm-$V) = 0.14
$\pm$ 0.04 \citep{bragaglia}, for a cluster metallicity of [Fe/H]
= +0.09 $\pm$ 0.03.

    In the next section we briefly discuss the observations
that have been obtained for this study.  This is followed by a
short discussion of the reduction procedures (a more complete
discussion is given in Paper I).  The science goals are then
addressed one at a time by first determining the center and radial
extent of the cluster which is a key for accurate star counts.  We
then discuss star/galaxy classification, present our CMD and
derive the distance to the cluster. This distance and the
reddening are used to fit a new generation of isochrones
(calculated especially for this project) and determine the cluster
age. Next we present luminosity functions, mass functions, and
binary star analysis for NGC 6819 which provide insight into the
dynamical evolution of the cluster. Finally, we provide a detailed
analysis of the white dwarf candidates in NGC 6819 which raise
further questions and motivate spectroscopic classification.

\section{Observations and Reductions}\label{observations}

    The observational data for NGC 6819 comes from the
first night of a three night observing run in October 1999, using
the CFH12K mosaic CCD on the Canada-France-Hawaii Telescope. This
CCD contains 12 2048 $\times$ 4096 pixel (0$\farcs$206) chips that
project to an area of 42$'$ $\times$ 28$'$ on the sky, which is
much larger than the cluster's radius, $\sim$9$\farcm$5.  We
obtained nine 300 second images in both B and V, as well as single
50 and 10 second images in each filter. At a later time, we also
acquired short 1 second frames so as to obtain unsaturated images
of the bright stars in the cluster. Blank field images are not
necessary as the outer four chips of the mosaic can be used to
correct for field star and galaxy contamination. The seeing for
the images deviated slightly from mean values of 0$\farcs$70 in V
and 0$\farcs$90 in B (see Table 1). Photometric skies combined
with this good seeing produced very sharp images of the cluster. A
true color image created from the processed V, B and R images is
shown in Figure \ref{mosaic}.

\notetoeditor{I would appreciate this colour figure (Figure
\ref{mosaic}) being placed alone on one page.  The size of the
final printed figure should be slightly smaller than as is when
you load the file.  Please use xv on Kalirai2.fig1.tif for expected
image quality.}

    We also obtain several flat-field, bias, and dark frames in
order to pre-process the data.  We chose to process the data using
the FITS Large Images Processing Software (FLIPS), which was
developed by Jean-Charles Cuillandre at CFHT (Cuillandre 2001).
FLIPS is a highly automated package which performs similar tasks
to the \sc iraf \rm option \bf mscred\rm.  A more detailed
discussion of the processing using FLIPS can be found in \S 3.1.1
of Paper I. Additionally, the methods employed within FLIPS to
average the nine images in V and B are discussed in \S 3.1.2 of
the same paper. These individual images were dithered from one
another to prevent stars from landing on bad pixels in more than
one image.

    The photometric calibration was obtained from numerous
exposures of the Landolt standard field SA-92 \citep{landolt}.
Detailed methods used to calibrate the data set are discussed in
\S\S 5.1 and 5.2 of Paper I.  Table 2 of Paper I summarizes the
number of exposures that were obtained in each filter and exposure
time for SA-92, and Table 3 summarizes the number of stars (of
varying air-mass) that were used on each CCD to calibrate the
data.   The photometric uncertainty in the zero points for the
standard stars during this night was measured to be $\sim$0.015 in
V and $\sim$0.014 in B. The air-mass coefficients were determined
to be 0.088 $\pm$ 0.01 in V and 0.165 $\pm$ 0.005 in B, both in
good agreement with CFHT estimations of 0.10 and 0.17
respectively. The color terms were averaged over the three night
observing run and are in agreement with CFHT estimations in the V
filter and slightly lower than estimations for the B filter.

    The data was reduced using a preliminary version of the new
TERAPIX photometry routine PSFex (Point Spread Function Extractor)
(E. Bertin 2000, private communication). This program is a new
part of SExtractor \citep{bertin}, which is commonly used for
star/galaxy classification.  We use a separate, variable PSF for
each CCD in the mosaic.  Further information on PSFex can be found
in \S 4 of Paper I.

\section{Star Counts and Cluster Extent}\label{starcounts}

    The existing measures of the apparent diameter of NGC 6819 makes
it one of the smallest, rich open star clusters known.  No clear
definition of the methods used to determine the previously adopted
cluster size have been published, however these results were
derived using much smaller CCDs or photographic/photoelectric
instruments which do not include a significant estimation of the
background stars directly around the cluster. Most early estimates
favored a larger cluster radius of 7$'$ (Barkhatova, 1963) or
6.5$'$ (King, 1964; Burkhead, 1971), although some are as small as
4$'$ (Lindoff, 1972).  Each of the CFH12K CCDs are 7$'$ in their
shortest direction, and therefore the entire cluster would be
concentrated within the inner 4 of the 12 CCDs if we center the
cluster in the middle of the mosaic.  To determine the cluster
center we first count stars in thin vertical and horizontal strips
across the mosaic.  This method is valid since there is no obvious
trend in the background distribution, therefore suggesting no
biasses from differential extinction across the CCD's.  The
resulting profile was less than gaussian in shape, and produced
limitations on the accuracy with which the center could be
determined.  An alternative approach which we implemented
consisted of estimating the center, and then counting the number
of stars in four equal quadrants around this center. Next we moved
the center location around a small area and re-counted the number
of stars until the value agreed in all four quadrants. Both
approaches lead to similar values which estimate the center of the
cluster to be at (x,y) = (6250,4020) on a global coordinate system
($\alpha_{J2000} = 19^{h}41^{m}17.7^{s}, \delta _{J2000} =
+40^{o}11'17''$). The error in each of the x and y directions is
$\sim$40 pixels (8$\farcs$2). This system combines the 6 chips
(horizontal) and 2 chips (vertical) on the mosaic (2048 pixels
horizontal/CCD $\Rightarrow$ 12288 total and 4096 pixels
vertical/CCD $\Rightarrow$ 8192 total) into one coordinate system
taking into account the gaps between the CCDs. This estimate
places the center of the cluster in the top right hand corner of
chip 02 (see Figure 2).

    The extent of NGC 6819 can now be found by counting the number
of stars in different annuli around the center of the cluster (see
Figure \ref{ccdann}).  To avoid significant biasses from selection
and incompleteness effects, we only use the stars with 15 $\leq$ V
$\leq$ 20.  Each successive annulus that we use is 1$'$ in width,
with the exception of those near the center of the cluster (see
Table 2 for annulus geometry). The value of the number of stars in
each annulus is then normalized by the area of the respective
annulus. We expect a drop off and stabilization in the resulting
distribution of number of stars vs. radius from the center as soon
as we clear the cluster and are simply counting the constant
background.  This approach is the first ever for this cluster and
is only made possible due to the large size of CFH12K. Figure
\ref{extentfig} shows the results for the extent of NGC 6819, and
indicates the cluster to be larger than all previous estimates. We
find a drop off caused by the boundary of the cluster and the
background between R = 8$\farcm$0-9$\farcm$5 (1$'$ = 290 pixels).
There are most likely still a very small number of cluster member
stars outside of R = 9$\farcm$5, however it is difficult to
resolve these from background.

\subsection{Comparison to King Model}\label{kingmodel}

    The mass range of the stars on the main-sequence between the two
magnitude cuts given above is not very large (0.70 - 1.5
M$_\odot$). In a classic series of papers King (1962; 1966)
described how the density distributions of stars in globular
clusters, open clusters, and some elliptical galaxies can all be
represented by the same empirical law.  This single-mass density
law is very simple to use and involves only three parameters: a
numerical factor, a core radius, and a limiting (tidal) radius.
For systems with a large range in mass (more than a factor of 10),
there are also more complicated multi-mass King models that can be
used to represent density profiles with more relaxed constraints
on mass ratios and distributions (eg. DaCosta \& Freeman 1976). In
order to fit a King model to the density profile shown in Figure
\ref{extentfig}, we first subtract off the background star counts
from each annulus (described in \S \ref{incompleteness}). Next we
determine a value for the tidal radius of NGC 6819.  The tidal
radius of a star cluster could potentially cause the cluster
extent to be truncated at a finite value.  This is determined by
the tidal influence of massive objects in the Milky Way (eg.
GMCs), which will remove the highest velocity stars from the
cluster as they venture out to large distances from the center.
The tidal radius for NGC 6819 can be estimated using equation
(\ref{eqntidalradius}),

%-----------------------------
\begin{equation}
r_{t} \sim (\frac{m}{3M})^{1/3}D, \label{eqntidalradius}
\end{equation}
\medskip
%-----------------------------

\noindent where m is the mass of the cluster ($\sim$2600M$_\odot$)
(see \S \ref{massfunc}), M is the mass of the Galaxy within the
cluster orbit, and D is the Galactocentric distance
\citep{clemens} of NGC 6819 (8.17 kpc).  Therefore, the tidal
radius for NGC 6819 is \it r\normalfont$_{t}\sim$17 pc, much
larger than the cluster radius ($\sim$6.9 pc). Additionally,
\cite{wielen} showed that it is not expected that a cluster of the
size and mass of NGC 6819 would be dissolved before a minimum age
of $\sim$7 Gyrs (much older than the cluster age). This analysis
takes into account both the effects of the Galactic tidal field
for internal and external processes of cluster dissolution, and
the evaporation of a star cluster due to its internal relaxation
(see \S \ref{relax}). With an estimate of the tidal radius of NGC
6819, we then fit a simple King model to the data and adjust the
value of \it r\normalfont$_{c}$ until the best fit is obtained.
\it r\normalfont$_{c}$ is strictly a model parameter in a King
profile and has no physical meaning, although for many clusters it
corresponds to the radius at which the surface brightness drops to
half the central value. The best fit obtained is at \it
r\normalfont$_{c}$ = 1.75 pc, and this profile is shown in Figure
\ref{kingtest} to agree very well with the data.

\notetoeditor{I would appreciate it if Figure \ref{extentfig} and
Figure \ref{kingtest} (the two cluster extent figures) could be
placed side by side in the journal.}

\section{Stellarity} \label{stellarity}

    Source classification is very important for studies of faint
objects in the cluster so that an accurate distinction between
star or galaxy can be made.  The star/galaxy cut will affect both
evolutionary tests of stars that become white dwarfs and the
luminosity function for these stars. Our data set extends to V
$\sim$ 25 implying that most of the objects measured here are
faint and have low signal to noise ratios. We use SExtractor
\citep{bertin} to assign a stellarity index to all objects on all
CCDs in our data. This stellarity index is determined through a
robust procedure that uses a neural network approach.  The
coefficients of this neural network were found by Emmanuel Bertin
by training the routine in simulations of artificial data. The
classification scheme attempts to determine the best hypersurface
for an object which can be described by either of two different
sets of parameter vectors; one for stars and the other for
galaxies.

    Figure \ref{stellfig} shows the variation of this stellarity index
with magnitude. The difficulty arises in choosing a confidence
limit to separate out stars from galaxies.  Objects with an index
of 0 are most likely galaxies and objects with an index of 1
appear stellar. Previous studies not involving faint objects have
often adopted a very high, 0.95, cutoff (eg. von Hippel and
Sarajedini 1998). However, some of the objects in the `clump' seen
in Figure \ref{stellfig} at 23 $\leq$ B $\leq$ 24.5 and 0.75
$\leq$ stellarity B $\leq$ 0.95 are most likely stars.  We find a
clear separation between a white dwarf cooling sequence and
background objects at approximately a 0.50 stellarity limit, and
an even better separation at 0.75, however at this more strict cut
we lose some faint objects (see \S \ref{wdanalysis}).  Later, we
will use a statistical method to eliminate possible galaxies and
background objects.  We note however that it is difficult to
determine this cut accurately without a spectroscopic
classification for these objects.

    We can estimate the number of galaxies that we expect in our
cluster field (inner 9$\farcm$5), by considering galaxy counts
\citep{woods} at high latitude and correcting for extinction in
our field. Based on these statistics and prior to any stellarity
cut, there are far more objects in our faint magnitude bins than
the number of expected galaxies. For example, we expect less than
7\% of all objects to be galaxies for 21 $\leq$ V $\leq$ 22, less
than 22\% for 22 $\leq$ V $\leq$ 23, whereas the number of
expected galaxies rises to $\sim$70\% for 23 $\leq$ V $\leq$ 24.
Clearly, this last bin is of interest as the number of expected
galaxies is comparable to the total number of all objects.
However, after applying a 0.75 stellarity cut, we eliminate 1673
of 1701 objects from this bin, a number far greater than the
number of expected galaxies, 1189. Therefore we can conclude that
a significant number of faint objects in the cluster field are
most likely stellar, however we may need to relax our 0.75
stellarity cut to include some of these fainter stars. A cut at
0.50 provides a better agreement between the number of objects
thrown out and the number of expected galaxies. Further details of
the stellarity index of faint blue objects in NGC 6819 is given in
\S \ref{whitedwarf}.

\section{The NGC 6819 Color-Magnitude Diagram}\label{cmd}

Figure \ref{clusterblank} exhibits CMDs for both the entire
cluster (R $\leq$ 9$\farcm$5) and a blank field taken from an
equal area of the outer chips on the mosaic. The cluster CMD
(left) shows a very tight main-sequence and turn-off region. For
this diagram we adopt a 0.50 stellarity confidence limit.  The
short 1 second exposures have allowed for the measurement of a red
giant clump and a giant branch at the bright red end of the CMD. A
significant contribution of these red giant stars is rare in most
open clusters due to their young age and poor population, but are
crucial for testing stellar models (see \S \ref{theory}). Some
potential white dwarf candidates can also be seen in the faint
blue end of the CMD. Further analysis of these stars is presented
in \S \ref{whitedwarf} where we fit cooling curves to the white
dwarf sequence after eliminating field objects and applying more
stringent confidence limits.  We also note the presence of a
significant contribution from approximately equal mass binary
stars in the cluster.  This sequence, which lies slightly above
and parallel to the main-sequence, is discussed in \S
\ref{binaries}.

\subsection{Cluster Distance by Main-Sequence Fitting}\label{msfitting}

    NGC 6819 is four times the age of the Hyades \citep{perryman}
and therefore a much smaller region of the CMD is available to fit
to the Hyades main-sequence fiducial for the cluster distance
because the high mass stars have evolved off the main-sequence.
Fortunately, our deep CCD photometry provides a longer un-evolved
main-sequence for the fitting.  Distance determination by
main-sequence fitting is also highly dependent on the adopted
reddening value of the cluster. For a reddening of E(B$\rm-$V) =
0.10, we find (m$\rm-$M)$_{V}$ = 12.30 $\pm$ 0.12. Correcting for
extinction (A$_{V}$ = 3.1E(B$\rm-$V)), the absolute distance to
NGC 6819 is therefore 2500 pc.  Our value is slightly larger than
most previous studies due to the lower adopted reddening value:
Rosvick and VandenBerg (1998) (d = 2350 pc), Auner (1973) (d =
2170 pc), and Lindoff (1972) (d = 2200 pc).

\subsection{Theoretical Isochrones}\label{theory}

    Testing theoretical stellar evolutionary models requires knowledge
of the cluster metallicity, reddening, and distance.  This testing
is very important in order to further refine the models.  For
example, theoretical models involving the amount of convective
core overshooting that should be used or the temperature at which
the slope changes at the faint end occur (caused by opacity
effects), lack observational constraints.  Additionally, the
testing allows for an age determination for the cluster. The
present photometry of NGC 6819 is ideal for these comparisons as
there are a large number of stars on a tightly constrained
main-sequence. The turn-off of the cluster, as well as the red
giants and giant branch stars are also useful as they allow for
tests of stars in the helium burning phases of evolution. These
are important for model calibration.

    The tracks were built by adopting the ATON2.0 code for stellar
evolution, a detailed description of which can be found in Ventura
et al. (1998). Convection has been addressed within the Full
Spectrum of Turbulence (FST) framework (Canuto \& Mazzitelli
1992). Chemical mixing and nuclear burning have been treated
simultaneously with a diffusive scheme: convective velocities have
been computed according to eqs. (88), (89) and (90) in Canuto et
al. (1996). Core-overshooting in the models has been described by
means of an exponential decay of turbulent velocity out of the
formal convective borders as fixed by Schwarzschild's criterion;
this behavior of velocity is consistent with approximate solutions
of the Navier-Stokes equations (Xiong 1985), and with the results
of numerical simulations (Freytag et al. 1996).  A value of
$\zeta$ = 0.03 of the free parameter giving the e-folding distance
of the exponential decay has been adopted.  We adopt grey
atmospheres for all models of mass above 0.6 M$_\odot$.  For
masses M $\leq$ 0.6 M$_\odot$ the models of \cite{montalban},
which adopt boundary conditions derived from the non grey model
atmosphere grid NextGen \citep{hauschildt}, are employed.  The
isochrone has been calculated by interpolating in mass between the
stellar tracks, according to the scheme by Pols et al. (1998).
Finally, the transformations between the theoretical and the
observational plane have been accomplished by adopting the colors
of Bessel et al. (1998).

    Figure \ref{msisochrone} shows the fit of the NGC 6819 data to a
2.5 Gyr old stellar isochrone.  We use a reddening value of
E(B$\rm-$V) = 0.10 and a derived distance modulus of
(m$\rm-$M)$_{V}$ = 12.30 $\pm$ 0.12 for this and further
theoretical comparisons. It is clearly seen that the slope of the
bright main-sequence as well as the turn-off and red giant clump
are all in excellent agreement with this model, which uses Z =
0.02.  Although still quite good, the fit to the lower
main-sequence is slightly blue of the data, which could be a
result of the photometric calibration of the data or,
alternatively, the color transformations in the model.

\section{Luminosity Functions}\label{lumfunc}

    The CMD of NGC 6819 indicates some similarities to other old
open star clusters such as M67, NGC 6633, NGC 752 \citep{francic},
and NGC 188 \citep{vonHippel2} with regards to the distribution of
stars. For example, by simply looking at the density of stars
along the main-sequence as a function of magnitude, it is clear
that the majority of the cluster members are bright, not faint.
After accounting for incompleteness at the faint end, such an
effect can be investigated by accurately counting the number of
cluster stars as a function of magnitude.  This resulting
luminosity function for such a cluster will clearly be either flat
($\frac{dN}{dM_V} \sim$ 0) or slightly negatively sloped
($\frac{dN}{dM_V} \leq$ 0). We define the cluster stars by first
creating a main-sequence fiducial (clipping the color at
3.5$\sigma$ from the mean) after isolating the main-sequence from
the background distribution.  We then use a clipping routine to
create an envelope around this fiducial based on the errors of the
stars (envelope broadens out towards faint magnitudes).  The
counting of the stars is done within this envelope, for both the
cluster CMD and the background CMD (Figure \ref{clusterblank}),
with the cluster luminosity function coming from the difference
between the counts in the two fields. However, in order to
accurately count stars, we must first determine incompleteness
corrections.

\subsection{Incompleteness Corrections}\label{incompleteness}

    The combination of cluster, telescope, detector, and reduction
routines leads to an incompleteness factor in the number of
detected objects.  For bright stars, this factor is typically
negligible, however, for fainter stars (V $\geq$ 20) it becomes
increasingly more important to determine how many stars have been
missed in the study.  In order to better understand the
incompleteness of our data we produce an artificial catalogue of
input stars for which we know the magnitudes and colors.  We add a
small number of stars uniformly in 5 trials so as not to affect
the crowding statistics of the field, and eventually obtain a
distribution of stars that lie on a similar slope to the raw
luminosity function of the cluster.  Next, we reduce this data set
in an identical manner to the data in the cluster and measure the
number of stars per magnitude bin that were recovered (see Figure
\ref{incompfig}). An identical analysis is also carried out for
the background fields. The completeness corrected number of stars
in the cluster can then be determined in 3 steps. First, we
multiply the cluster field incompleteness correction by the number
of stars in the cluster. Second, we multiply the blank field
incompleteness correction by the blank field stars. Finally, we
subtract the two and obtain the corrected star counts.  In this
analysis the cluster field stars are simply the stars for the
respective annuli in the inner 9$\farcm$5 of the center of the
cluster. For the blank field star counts, we use all of the stars
in CCDs 00, 06 and 11 (ie. three of the outer four in Figure
\ref{ccdann}) as well as the outer parts of the intermediate CCDs
(01, 04, 07 and 10), and then scale the numbers to match the area
of the corresponding inner cluster annuli.  This approach works
well as we have shown that the cluster extent (9$\farcm$5) does
not extend half way into the intermediate CCDs (01, 04, 07 and 10)
(see Figure \ref{extentfig}).  CCD 05 is ignored in the background
as it has a very low sky background level and we found some
abnormal zero point shifts and other errors throughout the
analysis.

    As expected we find that the current data set is quite complete at
the bright to intermediate magnitudes, however it becomes
increasingly more incomplete towards the faint end (60\% complete
at V $\sim$ 22). Incompleteness variations due to annuli are very
small as the crowding is not a significant concern for these
fields. However, we note that the incompleteness corrections were
slightly higher for faint magnitudes in chip 02, which is also the
CCD in which the center of the cluster resides. This is most
likely due to crowding and scattered light from the slight
over-density of bright stars. The incompleteness corrections for
bright stars are almost identical for the inner vs outer annuli.

    For the white dwarf sequence in NGC 6819, we determine an
incompleteness correction based on a set of artificial stars
distributed at this location in the CMD (21 $\leq$ V $\leq$ 25,
-0.25 $\leq$ B$\rm-$V $\leq$ 0.4).  Here, we choose a constant
function, which produces an equal number of objects per magnitude
bin (see Figure \ref{incompfig}).  This is a good method for the
white dwarfs in the cluster as our data set is not faint enough to
detect the end of the cooling sequence so there is negligible
piling-up of stars at faint magnitudes. As expected, the
incompleteness corrections for these stars are less (they are more
complete) than for the main-sequence stars at a given magnitude
because they are brighter in the B band for a given V magnitude.
These results are important for continuity arguments presented in
\S \ref{wdcontinuity}. Incompleteness findings for both the
main-sequence and the white dwarf sequence are summarized in Table
3.

\subsubsection{Incompleteness Errors}\label{errors}

    With these star counts, it is also important to have an
estimation of the errors.  Bolte (1989) gives a careful account in
determining incompleteness errors in the analysis of M30.  This
reasoning assumes that the counting uncertainties are derived from
a Poisson distribution, and that the artificial star count
uncertainties are derived from a binomial distribution.
Furthermore the errors in the incompleteness and the raw star
counts are assumed to be uncorrelated.  If we write the corrected
number of stars in any magnitude bin as \it n \rm =
$\frac{n_{obs}}{f}$, where \it n\rm$_{obs}$ is the raw counts in
the bin, and \it f \rm is the ratio of the number of recovered
artificial stars to the number added for each magnitude bin (\it f
\rm = $\frac{n_{recovered}}{n_{added}}$), then the variance in \it
n \rm is

%-----------------------------
\begin{equation}
\sigma_n^2 = \sigma_{n_{obs}}^2(\frac{\partial n}{\partial
n_{obs}})^2 + \sigma_f^2(\frac{\partial n}{\partial f})^2.
\label{eqnvariancen}
\end{equation}
\medskip
%-----------------------------

The variance in \it n\rm$_{obs}$ is simply $\sigma_{n_{obs}}^2 =
\it n\rm_{obs}$, and the variance in \it f \rm is $\sigma_f^2 =
\frac{f(1-f)}{n_{added}}$.  Performing the partial
differentiations in equation (\ref{eqnvariancen}) gives the
desired variance in \it n\rm,

%-----------------------------
\begin{equation}
\sigma_n^2 = \frac{n_{obs}}{f^2} +
\frac{(1-f)n_{obs}^2}{n_{added}f^3}. \label{eqnerrors}
\end{equation}
\medskip
%-----------------------------

We use this method of determining errors for both the field and
background stars, and then add the errors for the difference in
quadrature.

\notetoeditor{I think Table 4 looks better in landscape format}

    The final corrected star counts are presented in Table 4.
In this table, the first row of each magnitude bin consists of raw
counts (cluster field $\rm-$ blank field) whereas the row
underneath this one contains the incompleteness corrected numbers
(correction applied from Table 3).  Also shown in parentheses are
the errors in these counts, as calculated from the analysis given
above.  The corresponding global luminosity function is plotted in
Figure \ref{lumfuncfig}, where the dashed line represents the raw
counts and the solid line the incompleteness corrected counts. As
expected, the luminosity function is almost flat (slightly
negatively sloped), most likely due to dynamical evolution in this
relaxed cluster (see \S \ref{relax}).  We do not see a drop off at
the faint end of the luminosity function. Although large errors
make it difficult to determine, this evidence suggests that the
lowest mass main-sequence stars may not have been detected in this
deep photometry.  Integrating the luminosity function and
accounting for the evolved stars brighter than V = 15, gives a
lower limit to the total cluster population of $\sim$2900 stars.
This number makes NGC 6819 one of the richest open star clusters
known.

\subsection{Dynamical Relaxation}\label{relax}

    The cluster luminosity function is determined by the initial mass
function as well as the subsequent effects of dynamical evolution
to the present epoch.  The initial distribution of stars \it of
any mass \normalfont will roughly follow a density profile given
by a King model -- ie. an isothermal sphere \citep{binney}.
Therefore the initial density of stars for any mass will always be
highest in the center of the cluster and decrease as a function of
increasing radii from the center.  Equipartition of energy between
the stars of different masses occurs on a timescale characterized
by the relaxation time of the cluster.  If the cluster is older
than its relaxation time, then we can expect that the stellar
encounters within the cluster have caused the stars to relax
rapidly toward equipartition, with low mass stars travelling
faster than the high mass stars.  To estimate the relaxation time
for NGC 6819, we must first determine the crossing time of a star
across the cluster, \it t\rm$_{cross}$ = $\frac{R}{v}$.  Since we
know both the distance to the cluster (see \S \ref{msfitting}) and
its angular extent (see \S \ref{starcounts}), it is trivial to
solve for the linear radius ($\sim$6.9 pc). Additionally, the
velocity of a star across the cluster can be calculated using \it
v\rm$^{2} \sim \frac{GNm}{R}$, where \it Nm \rm is the mass of the
cluster (estimated in \S \ref{massfunc}).  The relaxation time is
then given by equation (\ref{eqnrelaxationtime}),

%-----------------------------
\begin{equation}
t_{relax} \sim t_{cross}\frac{N}{8lnN}, \label{eqnrelaxationtime}
\end{equation}
\medskip
%-----------------------------

\noindent where $\frac{N}{8lnN}$ is the number of crossings of a
star which are required for its velocity to change by an order of
itself \citep{binney}.  NGC 6819 is a very rich cluster, with
about 2900 stars (see \S \ref{errors}), which gives a relatively
large relaxation time compared to average open clusters with
several hundreds of stars. However, this relaxation time (220
Myrs) is still a factor of 10 smaller than the cluster age (2.5
Gyrs), so we expect the cluster to be relaxed: the lower mass
stars gain energy through gravitational encounters with higher
mass cluster stars (which sink to the center of the cluster) and
slowly diffuse out of the cluster if their escape velocity is
great enough \citep{hawley}. Therefore, for NGC 6819, the lower
mass stars are more likely to be observed at a larger radii than
the high mass stars. The evolution does however depend on
parameters such as binary fraction and cluster richness
\citep{delafuentemarcos}.

\subsection{Mass Segregation}\label{masssegregation}

    Mass segregation and evaporation of low mass stars have been
suggested to occur in dynamically relaxed clusters since van den
Bergh \& Sher (1960) demonstrated from their photographic data
that many clusters had luminosity functions that apparently turned
over at faint magnitudes, unlike that seen for the field stars.
Subsequent observational studies suggested that the mass functions
for some open clusters have changed over time due to dynamical
evolution (Francic 1989; von Hippel \& Sarajedini 1998; Raboud \&
Mermilliod 1998; Hawley, Tourtellot \& Reid 1999). Some of these
clusters are quite young, such as the Pleiades, but other such as
NGC 2420 are as old as NGC 6819. There have also been some studies
which show that intermediate aged clusters do not suffer from mass
segregation. Sagar \& Griffiths (1997) looked at mass segregation
effects for five, distant open clusters and found that the effects
are not correlated with cluster age. The major difference between
the current study and these others, is that NGC 6819 is far richer
in stellar content than most of these clusters and this affects
the evolutionary scenario because the escape velocity increases
with the cluster mass.

    In order to look for evidence of mass segregation in NGC 6819,
we produce three luminosity functions for different annuli from
the center of the cluster.  To keep the statistics reasonable we
split the cluster into three components, a central portion
(0$'$-2$\farcm$5), middle portion (2$\farcm$5-5$\farcm$5), and
outer portion (5$\farcm$5-8$\farcm$5).  Figure \ref{masssegfig}
shows the luminosity function for each of these portions.  We have
normalized the luminosity function for each of the annuli with
respect to the first annulus at V = 17.  The shapes of the
luminosity functions provide clear evidence for mass segregation
in NGC 6819. The high mass stars on the main-sequence are clearly
concentrated in the central regions of the cluster, whereas the
outer annuli show a greater relative concentration of low mass
stars.

    In Figure \ref{6cmds} we plot six CMDs, each for an increasing
radial annulus to show the richness of the main-sequence as a
function of radial position. Clearly, the CMD for annulus A11
(8$\farcm$5 $\leq$ R $\leq$ 9$\farcm$5) contains very few
main-sequence stars, which is consistent with Figure
\ref{extentfig}. The general trend from inner to outer annuli in
this figure confirms that the faint stars are not centrally
concentrated. Surprisingly, Figure \ref{6cmds} also shows a
prominent binary sequence in the intermediate-outer regions of the
cluster (see Annulus A6 $\rightarrow$ 3$\farcm$5 $\leq$ R $\leq$
4$\farcm$5). Binaries are discussed in \S \ref{binaries}.

\notetoeditor{I would appreciate this figure (Figure \ref{6cmds})
being placed on atleast 2/3 of a full page}

\section{Mass Functions}\label{massfunc}

    The mass function of a stellar population is typically expressed
as the number of stars / unit mass / unit solid angle.  Since it
is the luminosity that is measured, not the mass, this mass
function is usually expressed as in equation
(\ref{massfuncequation}),
%-----------------------------
\begin{equation}
N(m) = N(M_V)[\frac{dM_V}{dm}]. \label{massfuncequation}
\end{equation}
\medskip
%-----------------------------
In this equation, \it m \rm is the mass, \it N(M\rm$_V)$ is the
luminosity function, and $\frac{dM_V}{dm}$ is the mass-luminosity
relation. Therefore, the observed luminosity function must be
multiplied by the slope of the M$_{V}$-mass relation to obtain the
mass function.  Typically, the mass function is assumed to be a
power law so that

%-----------------------------
\begin{equation}
\Psi(m) \propto m^{-(1+x)}, \label{salpeter}
\end{equation}
%-----------------------------
where x takes on a value of 1.35 in the work of Salpeter (1955). A
discussion of the observational constraints and differing values
of this slope in young clusters, young field stars, old open
clusters, low mass disk stars, globular clusters, and the Galactic
spheroid and halo is given by \cite{richer3}.  Additionally,
Francic (1989) has shown that the mass functions for some old
Galactic clusters (NGC 6633, NGC 752, and M67) are weighted
towards the higher mass stars. This analysis also showed the slope
of the mass function for younger open clusters to be about x = 1.
The inverted mass function for the older clusters may be due to
dynamical processes in the cluster which work to better retain the
higher mass stars.  We have shown that it is likely that these
processes may have already occurred in NGC 6819 (see \S
\ref{relax}). In order to better quantify the dynamical evolution,
we can observe the change in the mass function for different
annuli from the center of the cluster.

\notetoeditor{I would appreciate this figure (Figure
\ref{massfuncfig}) being placed on atleast 2/3 of a full page}

    We use the Rome theoretical isochrones (see \S \ref{theory})
to create a mass-luminosity relationship for NGC 6819.  The slope
of this mass-luminosity relationship is used to convert the number
of stars in each magnitude bin to the number of stars per unit
mass. We derive the slope using the end points of each magnitude
bin that is used in the analysis, and from this slope, derive the
mass function.  As expected, the global (R $\leq$ 9$\farcm$5) mass
function (bottom-right of Figure \ref{massfuncfig}) is almost flat
($\frac{dN}{dm}$ = constant). Fitting a power law to this global
function (as in equation (\ref{salpeter})) gives a value of x =
$\rm-$0.15.  For comparison, we also plot a Salpeter value (x =
1.35) in the global plot which is much steeper than the NGC 6819
mass function. Figure \ref{massfuncfig} also shows a series of
mass functions for annuli at increasing radial distances from the
cluster center. There is a systematic change in the slope of the
mass function with increasing radius (positive slope $\rightarrow$
negative slope) which is consistent with the expectations of
dynamical evolution in NGC 6819. Integrating the global mass
function provides a total cluster mass of $\sim$2600M$_\odot$.

\section{Binary Stars in NGC 6819}\label{binaries}

    The early stages of the dynamical evolution of star clusters
has been shown to be dominated by primordial binaries
\citep{heggie}.  These binaries can be explained by a number of
different formation mechanisms: capture events, fragmentation
processes during collapse of protostar, exchange events, and
orbital decay (see de la Fuente Marcos 1996; Hartigan et al. 1994;
Bodenheimer 1993).  The presence of binaries in a cluster is
expected to accelerate the cluster dispersion in the early stages
of cluster evolution.  Multicomponent cluster models indicate that
the early evolution will be accelerated for sparse clusters and
slowed down for very rich clusters \citep{delafuentemarcos2}. The
effects of these binaries on the cluster environment are estimated
to be less than the effects of stellar evolution.  A study of the
location of the binaries in NGC 6819 can be important for
constraints on future models as well as comparisons to binary
populations in younger clusters (Brandner et al. 1996; Ghez et al.
1993, 1994; Padgett et al. 1997).

    In order to quantify the location of the equal mass binaries
in NGC 6819, we create a fiducial that matches with the observed
binary population in annulus A6, and isolate the sequence from the
remaining stars on the CMD by eliminating objects that deviate by
more than 3$\sigma$ from the mean location of the fiducial. This
annulus produced the most prominent binary sequence (observing by
eye...see upper-right diagram in Figure \ref{6cmds}), however it
does not contain the highest fraction of binaries as the majority
of the main-sequence stars are also concentrated here.  The
fiducial is located between $\Delta$V = 0.72 and 0.77 above the
main-sequence fiducial depending on the magnitude. We then use
this fiducial to count the number of stars in a thin strip
centered on the fiducial in all annuli and the corresponding blank
fields for each. The results after subtracting the cluster fields
from the blanks are presented in Table 5.  A lower limit for the
global cluster equal mass binary content is found to be $\sim$
11\%, however the highest population is observed in the outer
regions of the cluster (A9 = 18\%, A10 = 25\%).  The errors in
these values are large (parenthesis in Table 5) and it is
difficult to speculate on any evolutionary effects of binaries
from this data, however we do note that the equal mass binaries of
NGC 6819 are more concentrated in the outer regions of the cluster
(with the exception of A8). This is not expected as the binaries
are higher mass and should sink to the center of a dynamically
relaxed cluster.  The reason for this may be that we are missing
the most massive binaries in this analysis.  The vertical
$\Delta$V = 0.76 magnitude shift caused by a system with a
mass-ratio of 1 (equal mass binaries) would cause the location of
the massive objects to coincide with the turn-off of the
main-sequence, therefore they are not counted in the present
analysis.

\section{White Dwarfs in NGC 6819}\label{whitedwarf}

    Studies involving old white dwarf stars in globular star clusters
are very important for obtaining a lower limit to the age of the
Universe.  At present this can only be done with the Hubble Space
Telescope as the end of the white dwarf cooling sequence for these
clusters occurs at extremely faint magnitudes (M$_{V}$ = 17.5 for
age = 12 Gyrs \citep{richer}). Determining ages in this manner
strongly depends on the theoretical models of cooling white
dwarfs. It has been recently shown that old hydrogen rich white
dwarfs may actually be much bluer than previously thought
\citep{hansen} due to atmospheric molecular hydrogen opacity
effects. These new models predict a strong blue `hook' in the CMD
for the cooling white dwarfs at an age of $\sim$8 Gyrs. Although
the white dwarfs in NGC 6819 are far too hot to form molecular
hydrogen which causes a suppression of flux in the infrared, they
are important for calibration and confirmation of the validity of
the models at brighter magnitudes that will be used in globular
cluster studies.

\subsection{Continuity Arguments and Field Object Subtraction}
\label{wdcontinuity}

    NGC 6819 is very rich in stellar content, thus a significant
number of both hydrogen (DA) and helium (DB) white dwarfs are
expected. We can predict the number of expected white dwarfs above
a limiting magnitude cut-off in the cluster by counting the number
of stars in the red giant phase and applying a continuity argument
for these stars.  From the masses of the red giant stars and the
Rome stellar evolutionary sequence, the lifetime of the red giants
in the `clump' (V = 13, B$\rm-$V = 1.2) is determined to be 5
$\times$ 10$^{7}$ years for models with no convective overshooting
and 9 $\times$ 10$^{7}$ years for an overshooting model. Clearly,
we favour the latter model as it agrees better with our data.
Additional evidence for the justification of core-overshooting
models for NGC 6819 was given in detail in the analysis of Rosvick
and VandenBerg (1998). The continuity argument which we apply to
determine the number of expected white dwarfs ($N_{WD}$) follows
from the hypothesis that all stars of mass less than $\sim$8
M$_\odot$ will evolve in to white dwarfs. First we determine the
number of objects in the red giant `clump' ($N_{RG}$) after
correcting for field star contamination; 13. Next we can use the
white dwarf cooling models to determine the white dwarf cooling
ages ($t_{cooling}$) at a certain magnitude (V = 23, 23.5, and 24)
that the stars have cooled to, for varying white dwarf masses (M =
0.5, 0.6, and 0.7 M$_\odot$). For post main-sequence evolution the
number of stars in a given evolutionary phase is proportional to
the time spent in that phase.  Therefore, we can estimate the
number of white dwarfs that we expect to see in the cluster and
compare this with the number observed after both field star
subtraction and incompleteness corrections (see \S
\ref{incompleteness}), by using equation (\ref{number}):

\medskip
%-----------------------------
\begin{equation}
N_{WD}(\leq M_{V}) = \frac{N_{RG}}{t_{RGB}}t_{cooling}(\leq
M_{V}). \label{number}
\end{equation}
\medskip
%-----------------------------

The field star subtraction is addressed statistically by comparing
the location of objects in the lower-left (faint-blue) section of
the cluster CMD to the background field CMD.  We take each object
within this location on the background CMD and eliminate the
corresponding closest object in the cluster CMD.

    We can estimate the uncertainties in the expected number of white
dwarfs in a similar manner to that used for the main-sequence
stars: use Poisson errors for both the observed number of white
dwarfs and red giants, and a binomial distribution for the
incompleteness errors (see \S \ref{errors}). Additionally, there
is an uncertainty in the white dwarf cooling age which is found by
multiplying the slope of the cooling curve by the error in the
magnitude as determined by PSFex.

    We find that for a model with a large amount of core-overshooting
(this produces a larger convective core $\rightarrow$ time in RG
phase increases) the predicted number of white dwarfs far exceeds
the number observed (to V = 24) at a strict 0.90 stellarity cut.
However, the predicted number agrees very well if we impose a less
stringent confidence limit of 0.80, especially for 0.7 M$_\odot$
objects. Similar results are also seen for intermediate
core-overshooting, where the number of predicted white dwarfs
agrees well with the number observed up to V = 23.5.  For a
brighter magnitude cut of V = 23, we find too many white dwarfs in
all but the low core-overshooting cases with M = 0.6 and 0.7
M$_\odot$. It is difficult to make predictions from this analysis
because the uncertainties remain large. The analysis however
indicates that a core-overshooting model is preferred for the
cluster if we are to consider all potential white dwarfs to our
limiting magnitude. This assumes that a substantial number
fraction of the white dwarfs are not tied up in binaries.  These
results are summarized in detail in Table 6, with errors in
parentheses.

\subsection{White Dwarf Analysis}
\label{wdanalysis}

\notetoeditor{Could Figure \ref{wdfigure1} be printed somewhat
larger than the normal figure size?}

Figure \ref{wdfigure1} shows the CMD for the cluster before any
source rejection (left) and after we have imposed some constraints
(middle and right). The constraints are (1) only accept objects
with a stellarity confidence index above 0.50 (middle), and (2)
only accept those objects which also survive a statistical
subtraction to remove field objects (right). Criterion (1) is
arbitrary as explained in \S \ref{stellarity}. Even at a 0.50
stellarity cut, some of these objects could still be faint
unresolved galaxies, AGN, or some other non-cluster object.  We
note however, that a significant portion of the remaining objects
are in fact stellar to within a 0.90 confidence. This is shown in
more detail in Figure \ref{wdfigure2}, where we have zoomed into
the hot faint end of the CMD.  Criterion (2) has been addressed by
eliminating possible cluster field objects that are in the same
vicinity of the CMD as background field objects as described in \S
\ref{wdcontinuity}.  We invoke this approach for a small region in
the CMD surrounding and including all possible cluster white
dwarfs (objects below the dashed line in Figure \ref{wdfigure1}
(middle)).  The statistical subtraction shows that there is an
over-density of objects in the cluster field, however, we can not
say for sure whether the objects that we have removed are in fact
not cluster objects.  In Figure \ref{wdfigure2} we also indicate
the stellarity for each of the white dwarf candidates, which is
found to be $\geq$ 0.90 for the majority of the objects above V =
24.

\subsection{Interpretation of Cooling Sequence}
\label{coolingsequence}

    The statistically subtracted and star/galaxy corrected CMD
(Figure \ref{wdfigure1} (right)) indicates a clear separation
between the white dwarfs and the field stars.  This potential
white dwarf cooling sequence is separated from these field stars
by an average of $\sim$0.6 magnitudes in color on the CMD. There
are very few objects between the two populations.  The putative
white dwarf cooling sequence revealed however is not particularly
tight as there is some evidence for a gap between this (the bluer
objects) and an redder potential white dwarf sequence
(apparent in Figure \ref{wdfigure2}). For the adopted reddening
value of the cluster, no reasonable mass white dwarf cooling
sequence fits the reddest objects in this location of the CMD.  It
is unlikely that photometric spread is causing some of these
objects (the redder objects) to deviate so much in color from
those that agree with the 0.7 M$_\odot$ cooling sequence
\citep{wood}. To better judge this, we have plotted a photometric
error bar as a function of magnitude at 0.5 magnitude intervals in
Figure \ref{wdfigure2}. These errors, as determined by PSFex, are
consistent with those found for the recovered artificially added
stars in the incompleteness tests. It is clear that the spread in
data points is larger than the error bars.  A more likely scenario
to explain the positions of these objects in the CMD is that they
are just excess background or foreground objects which were not
removed in the statistical subtraction.  Alternatively, some could
also be highly reddened background white dwarfs.

    For those objects which closely follow the 0.7 M$_\odot$ white dwarf
cooling sequence we determine a luminosity function after
correcting for both a 0.75 stellarity cut and an incompleteness
correction. The slope of this luminosity function agrees very well
with theoretical expectations \citep{fontaine} (see Figure
\ref{wdfinal}). The bright end of this theoretical function has
been slightly extended to include our brightest objects. There is
no obvious agreement with the theoretical function if all objects
in Figure \ref{wdfigure2} are considered. Although we believe the
bluest objects in this Figure to be bona-fide white dwarfs,
spectroscopic confirmation is required for these as well as those
that were eliminated in the statistical subtraction. Fortunately,
multi-object spectroscopy with instruments such as GMOS on Gemini
will allow measurement of multiple objects in the faint-blue end
of the CMD.

%-----------------------------

\section{Conclusion} \label{conclusion}

    We have obtained deep (V $\sim$ 25) CCD photometry of a 42$'$
$\times$ 28$'$ field centered on the old open star cluster, NGC
6819.  This photometry indicates both a larger cluster extent
($\sim$9$\farcm$5) and a much richer cluster population
($\sim$2900 stars, 2600 M$_\odot$, after incompleteness
corrections are applied) than previously estimated. Main-sequence
fitting of the un-evolved stars in the cluster indicates a true
distance modulus of (m$\rm-$M)$_{o}$ = 11.99 $\pm$ 0.18 for a
reddening value of E(B$\rm-$V) = 0.10. Isochrone fits with
up-to-date models are in excellent agreement with the data and
comparison with turn-off stars in the cluster provides an age
estimation of $\sim$2.5 Gyrs. Measurements of the luminosity
function and mass function of the cluster in concentric annuli
indicate the cluster to be dynamically evolved. Studies of the
cluster CMD suggest clear evidence for mass segregation in NGC
6819.  A lower limit of 11\% is found for the equal mass binary
fraction in the cluster, although this number is up to a factor of
two higher for some of the outer regions of the cluster. Analysis
of the faint blue section of the CMD indicate $\sim$21 high
probability WD candidates brighter than V = 23.5. Most of these
stars are scattered around a 0.7 M$_\odot$ white dwarf cooling
curve. Until we have spectroscopic confirmation of their WD
nature, it is premature to determine the initial-final mass
relationship for these stars.

\clearpage

%% No more than seven \figcaption commands are allowed per page,
%% so if you have more than seven captions, insert a \clearpage

%% There must be a \figcaption command for each legend. Key the text of the
%% legend and the optional \label in curly braces. If you wish, you may
%% include the name of the corresponding figure file in square brackets.
%% The label is for identification purposes only. It will not insert the
%% figures themselves into the document.
%% If you want to include your art in the paper, use \plotone.
%% Refer to the on-line documentation for details.

\clearpage

\figcaption[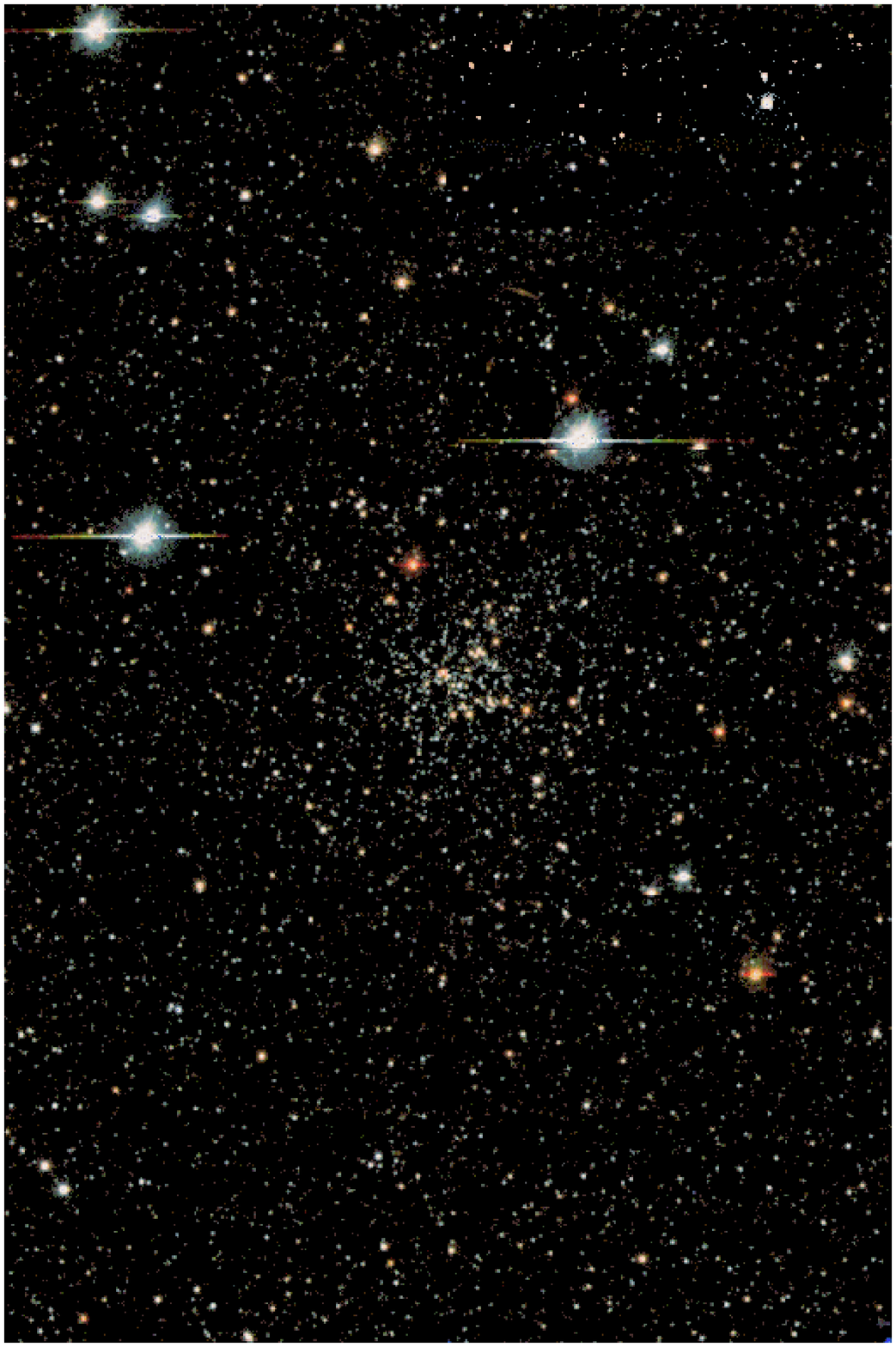]{True-color image from the individual
V, B and R 50 second frames.  The image size is 42$'$ $\times$
28$'$. \label{mosaic}}

\epsscale{0.75}

\plotone{Kalirai2.fig1.ps}

\clearpage

\figcaption[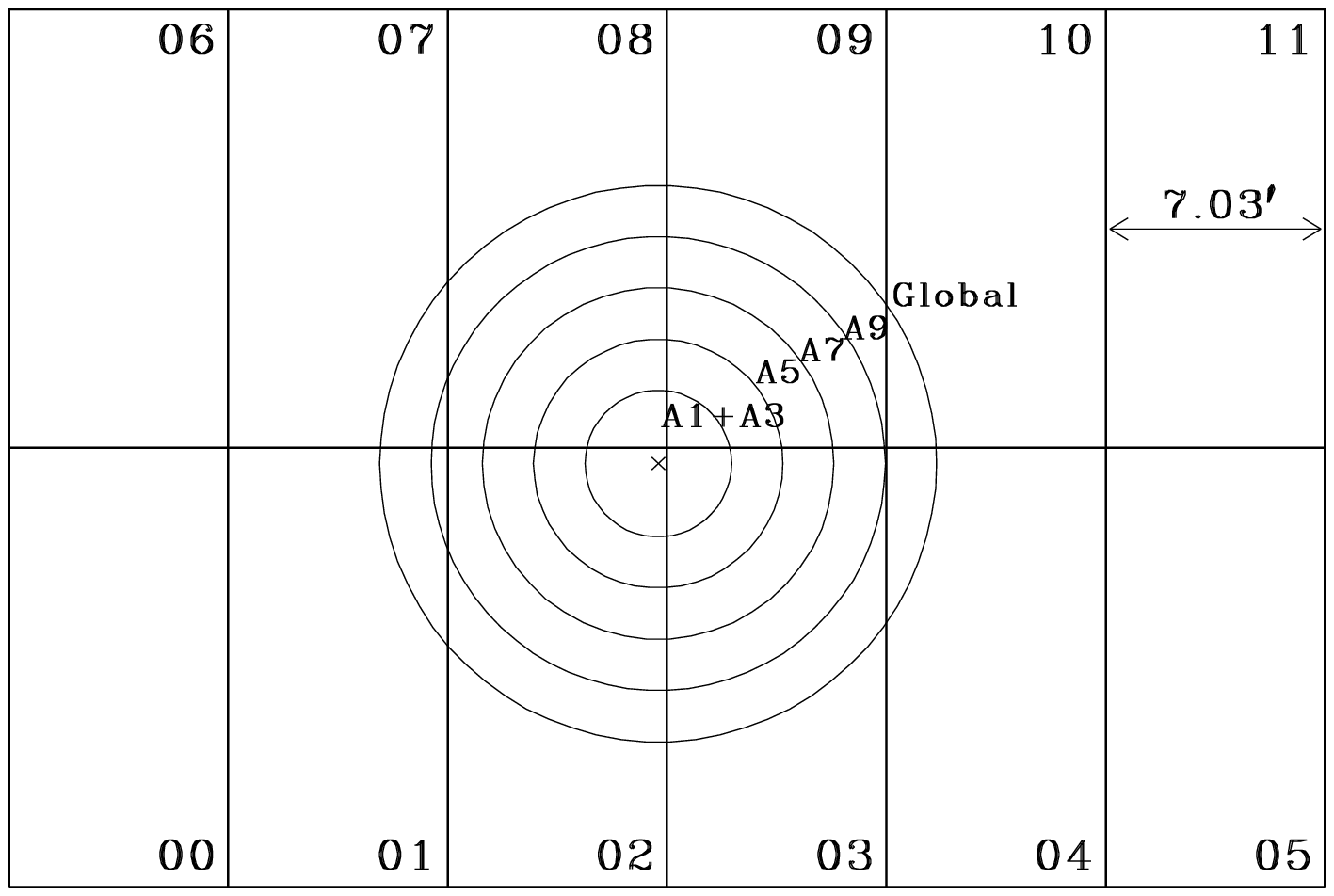]{Orientation of the 12 chips of
CFH12K are shown.  The small x in chip 02 marks the approximate
center of the cluster.  The outer radii of four annuli separated
by 2$'$ (1$\farcm$5, 3$\farcm$5, 5$\farcm$5, 7$\farcm$5 and
9$\farcm$5) are also shown.  \label{ccdann}}

\epsscale{1} \plotone{Kalirai2.fig2.eps}

\clearpage

\figcaption[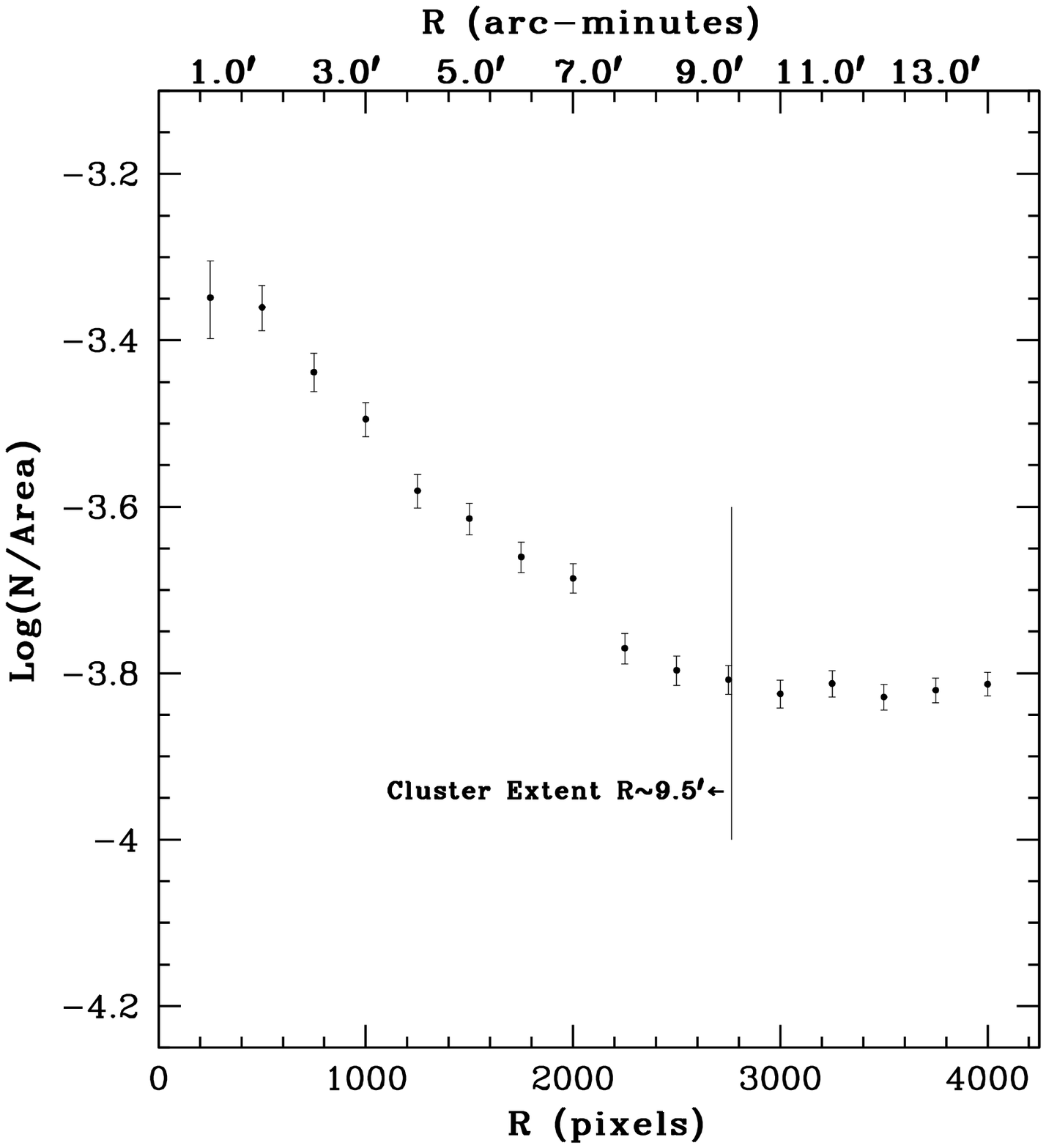]{Star counts in NGC 6819.  The
radius of the cluster is shown to be $\sim$9$\farcm$5. The flat
distribution of stars after this point is just background.  Error
bars reflect both Poisson errors in the cluster and blank field
star counts, and the error in locating the center of the cluster.
\label{extentfig}}

\plotone{Kalirai2.fig3.eps}

\clearpage

\figcaption[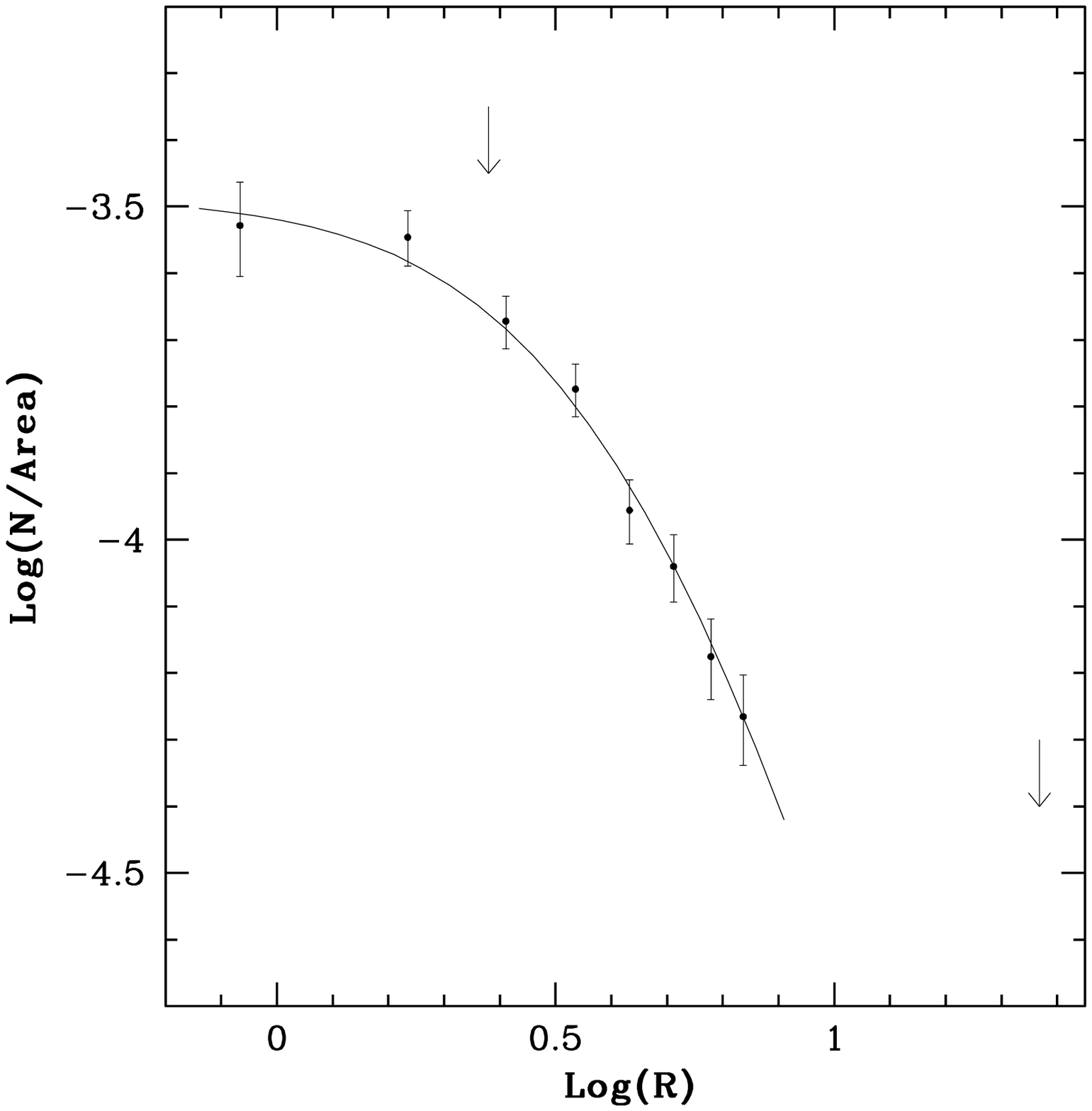]{Single-mass King model shown to
agree well with the cluster density distribution of NGC 6819. The
arrows correspond to the core and tidal radii of the cluster.
Error bars reflect both Poisson errors in the cluster and blank
field star counts, and the error in locating the center of the
cluster. \label{kingtest}}

\plotone{Kalirai2.fig4.eps}

\clearpage

\figcaption[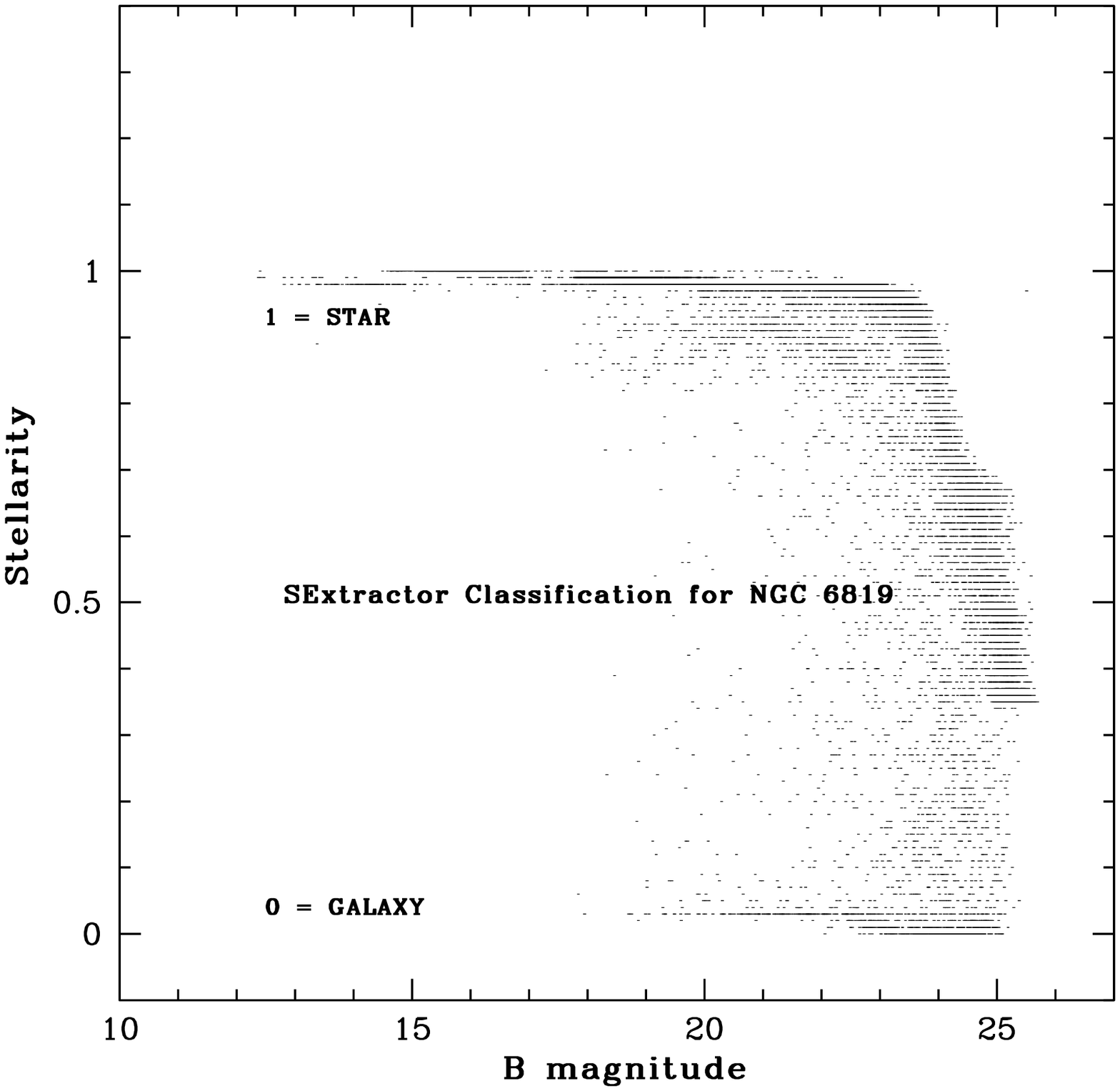]{Star/galaxy classification from
SExtractor indicating many sources that are most likely stellar
(stellarity = 1) as well as some that are most likely galaxies
(stellarity = 0).  The classification of the remaining objects is
addressed in \S\S 4 and 9.2. \label{stellfig}}

\plotone{Kalirai2.fig5.eps}

\clearpage

\figcaption[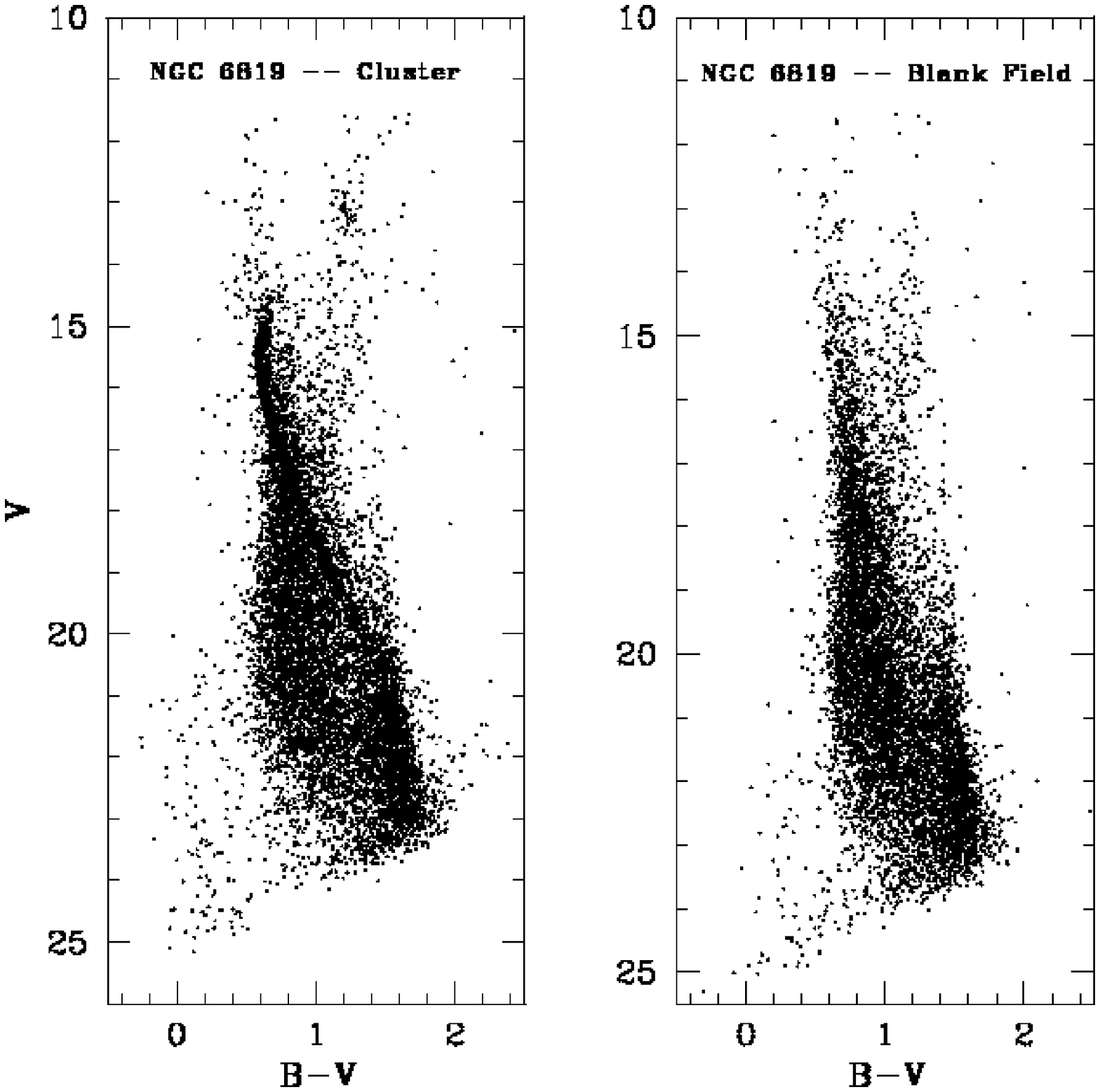]{Rich, tight main-sequence and
turn-off of NGC 6819, clearly seen (left). Also shown is a blank
field of equal area taken from the outer four CCDs of CFH12K
(right). A 0.50 stellarity cut has been applied to the data.
\label{clusterblank}}

\plotone{Kalirai2.fig6.eps}

\clearpage

\figcaption[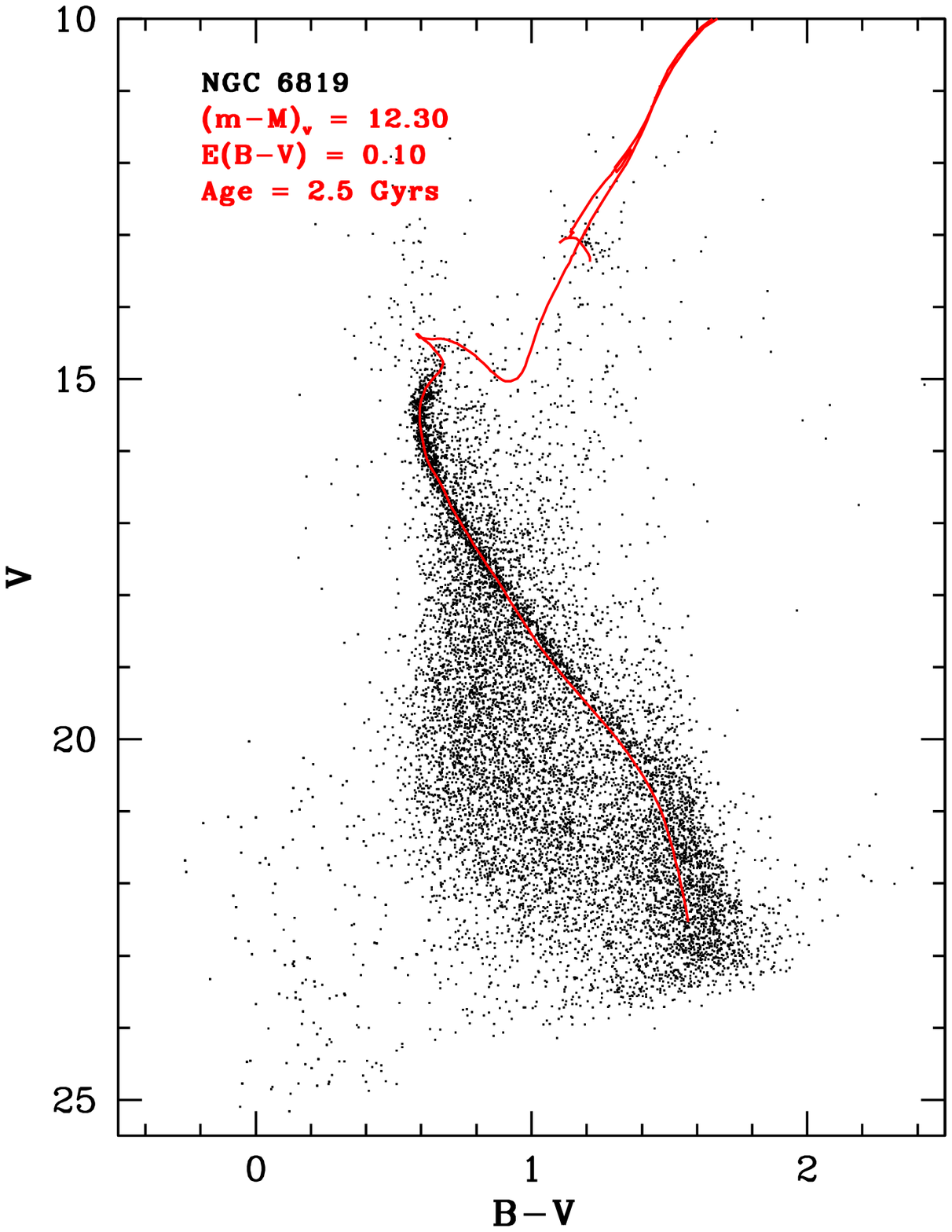]{Theoretical isochrone of age 2.5
Gyr, found to fit the turn-off well.  The slope of the
main-sequence and the location of the red giant clump also agrees
well with the isochrone.  Some potential white dwarfs are also
evident in the faint blue end of the CMD (we have applied a 0.50
stellarity cut). \label{msisochrone}}

\plotone{Kalirai2.fig7.eps}

\clearpage

\figcaption[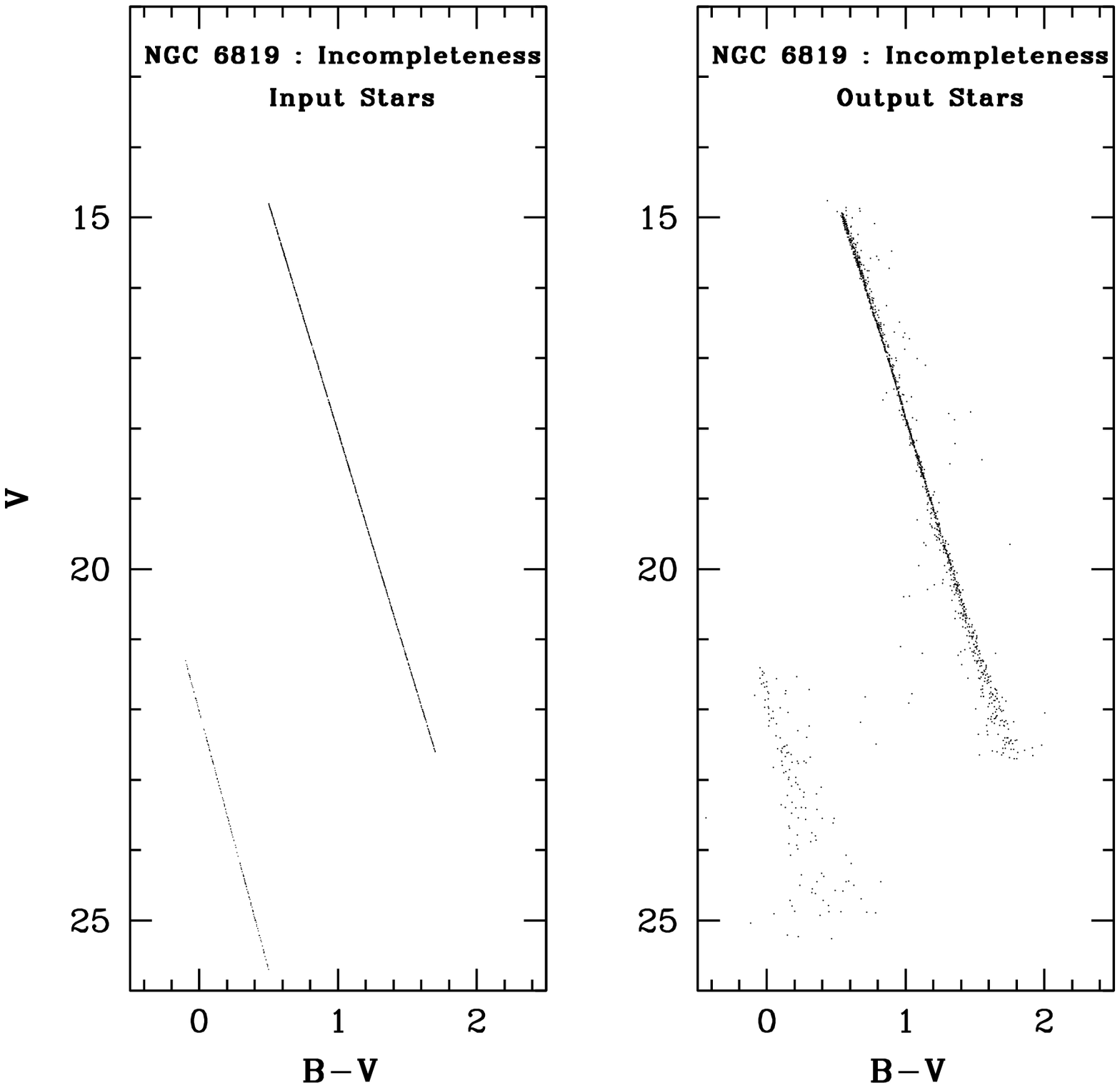]{Both input stars (left) and
recovered stars (right) in the incompleteness tests for both the
main-sequence and potential white dwarf cooling sequence.
\label{incompfig}}

\plotone{Kalirai2.fig8.eps}

\clearpage

\figcaption[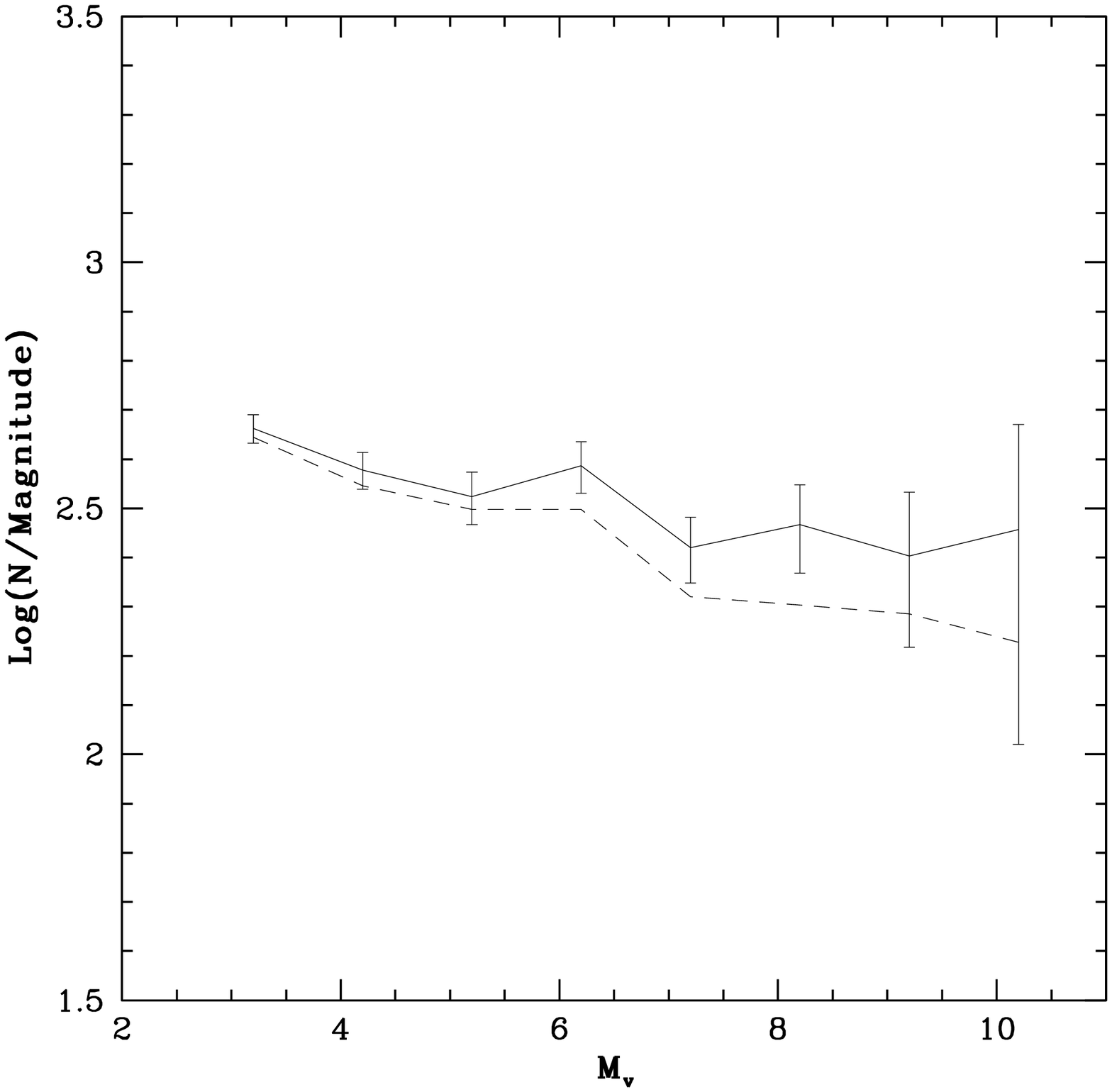]{Global (R $\leq$ 9$\farcm$5)
luminosity function shown before (dashed) and after (solid)
incompleteness corrections.  The almost flat luminosity function
is most likely due to dynamical evolution which has caused the
high mass stars to sink to the inner regions of the cluster.  The
error bars reflect a combination of Poisson errors and
incompleteness errors as discussed in \S 6.1.1 \label{lumfuncfig}}

\plotone{Kalirai2.fig9.eps}

\clearpage

\figcaption[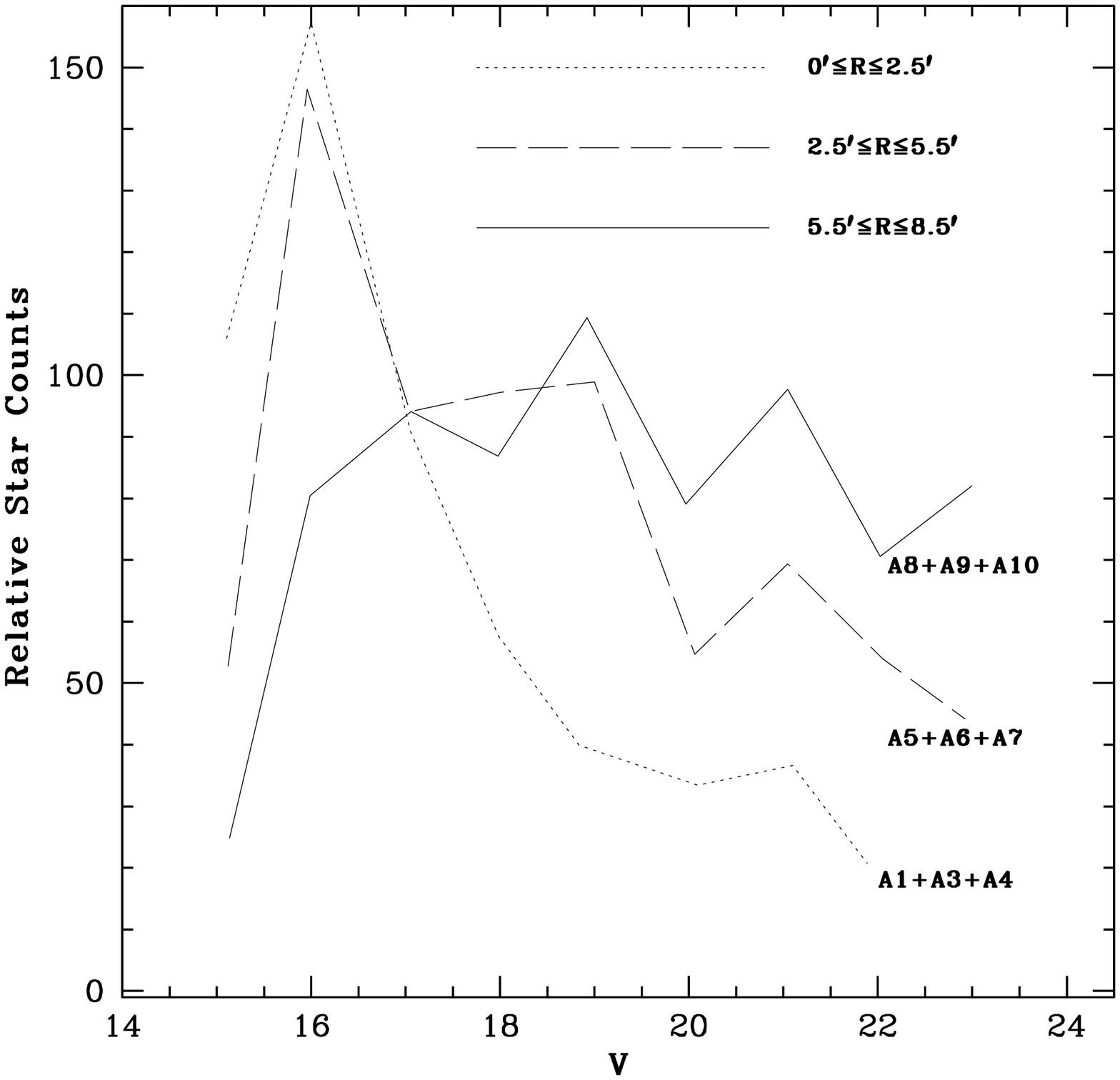]{Luminosity function shown for
three regions in the cluster.  The counts in the central and outer
annuli have been normalized to the number in the inner annulus at
V = 17. This demonstrates clear evidence for mass segregation in
NGC 6819. \label{masssegfig}}

\plotone{Kalirai2.fig10.eps}

\clearpage

\figcaption[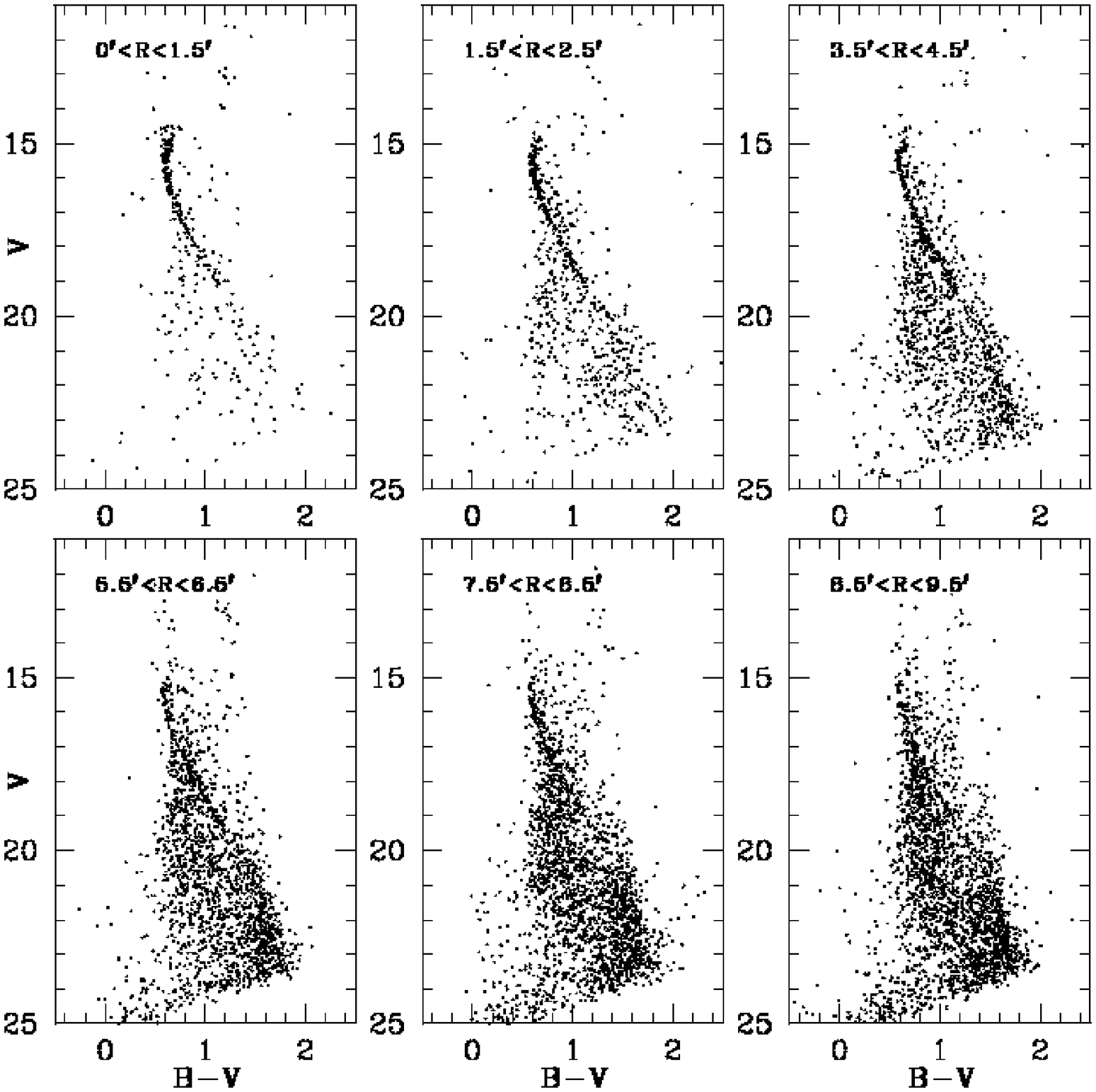]{CMD for NGC 6819 for each
alternate annuli, illustrating the main-sequence density as a
function of increasing radius from the center. The figure shows
that the lower mass (faint) stars are located in the outskirts of
the cluster.  A prominent binary sequence is also evident in the
intermediate regions of the cluster (top right). \label{6cmds}}

\plotone{Kalirai2.fig11.eps}

\clearpage

\figcaption[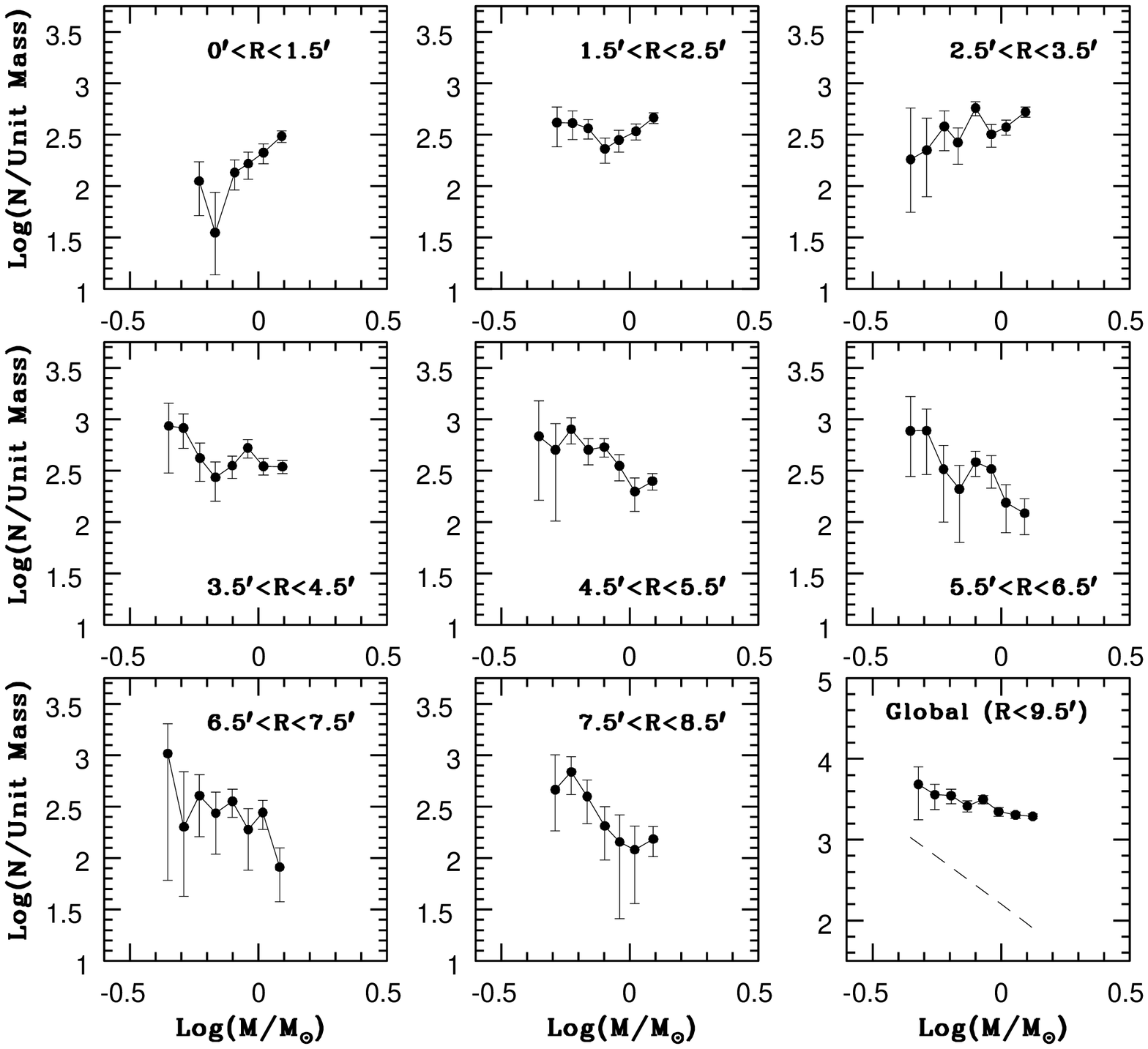]{Series of mass functions for eight
different annuli, illustrating a general trend of positive to
negative slope as a function of increasing distance from the
cluster center.  The global mass function (x = -0.15) of the
cluster is clearly flatter than a Salpeter value (x = 1.35, dashed
line, bottom right). The error bars are taken from the errors in
the luminosity functions (Poisson and incompleteness) and then
multiplied by the slope of the mass-luminosity relation.
\label{massfuncfig}}

\plotone{Kalirai2.fig12.eps}

\clearpage

\figcaption[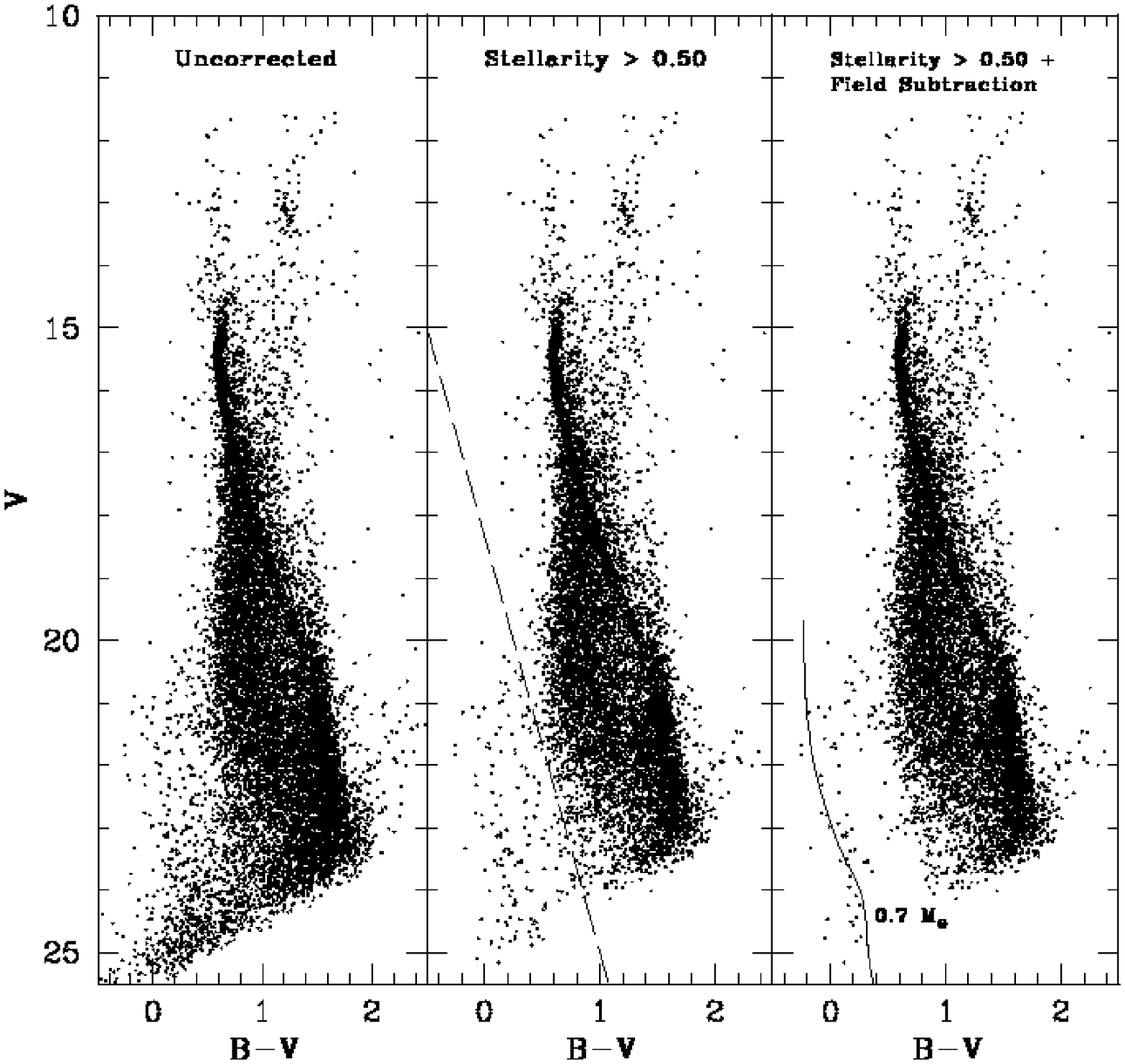]{Uncorrected CMD for NGC 6819
(left). There is a general spread of stars in the lower left
corner. After correcting for extended sources (center) and field
star subtraction (blue of dashed line, middle), a potential white
dwarf cooling sequence is evident (right). We also show a 0.7
M$_\odot$ white dwarf cooling sequence which agrees with the
bluest potential white dwarfs.  This analysis is a purely
statistical method of determining the most likely location on the
CMD of the cluster white dwarfs. \label{wdfigure1}}

\epsscale{0.95} \plotone{Kalirai2.fig13.eps}

\clearpage

\figcaption[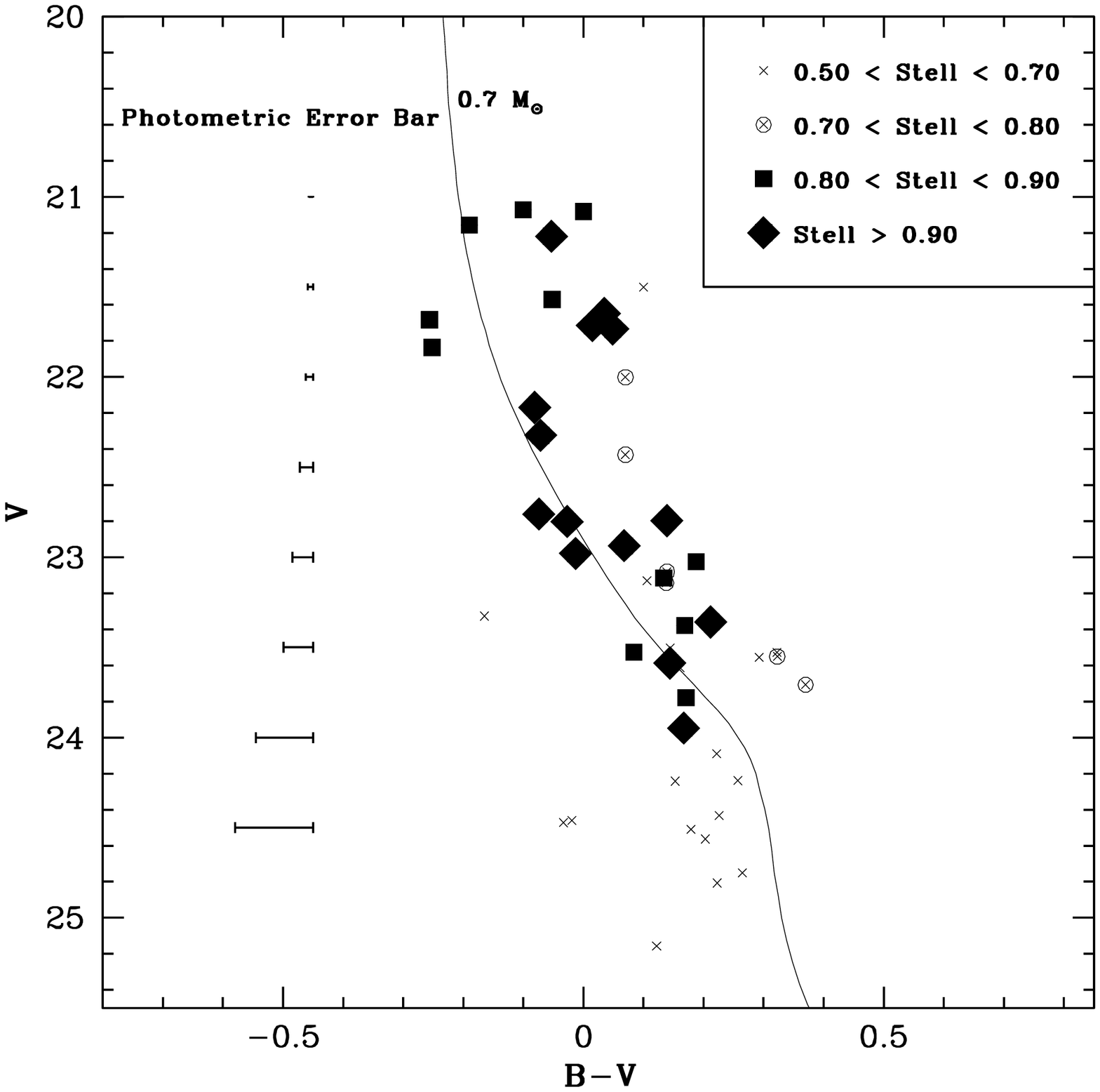]{All potential white dwarf
candidates within 0.3 magnitudes (color) of the 0.7 M$_\odot$
cooling sequence and after a statistical subtraction.  A
photometric error bar is also shown as a function of magnitude.  A
large number of the objects are determined to be very high
confidence stars (diamonds). \label{wdfigure2}}

\epsscale{1} \plotone{Kalirai2.fig14.eps}

\clearpage

\figcaption[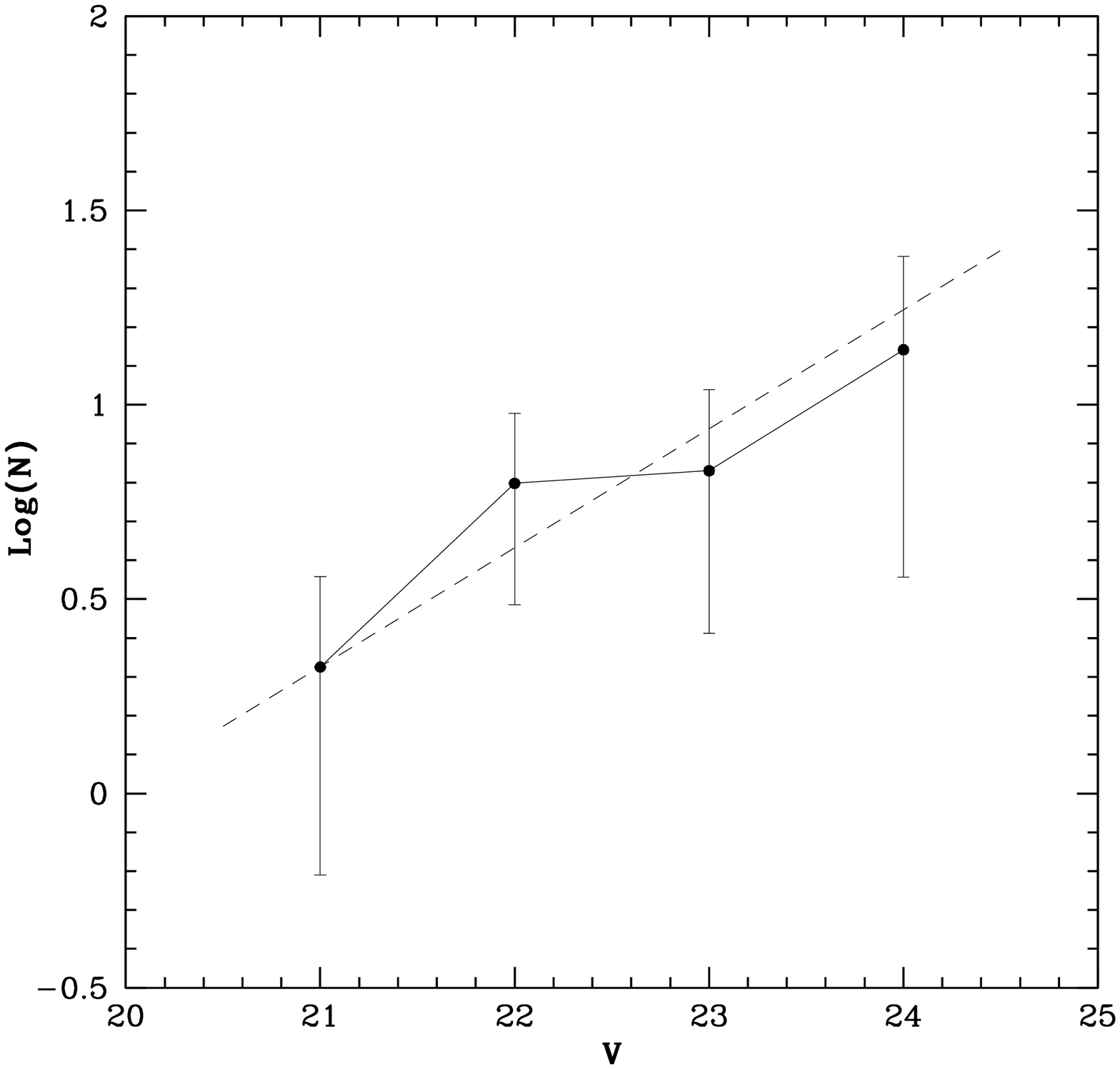]{Luminosity function of the
left-most trail of white dwarfs in the CMD; agrees well with the
slope of the theoretical luminosity function (dashed line--see \S
9.3).\label{wdfinal}}

\plotone{Kalirai2.fig15.eps}

\clearpage

%% Tables should be submitted one per page, so put a \clearpage before

%% Two options are available to the author for producing tables:  the
%% deluxetable environment provided by the AASTeX package or the LaTeX
%% table environment.  Use of deluxetable is preferred.
%%

%% Three table samples follow, two marked up in the deluxetable environment,
%% one marked up as a LaTeX table.

%% In this first example, note that the \tabletypesize{}
%% command has been used to reduce the font size of the table.
%% Note also that the \label command needs to be placed

\clearpage

\begin{deluxetable}{ccccc}
\tabletypesize{\scriptsize} \tablecaption{Observational Data for
NGC 6819 \label{table1}} \tablewidth{0pt}
\tablehead{\colhead{Filter} & \colhead{Exposure Time (s)} &
\colhead{No. of Images} & \colhead{Seeing ($''$)} &
\colhead{Airmass}} \startdata

V & 300 & 9 & 0.70 & 1.30 \\

 & 50 & 1 & 0.70 & 1.16 \\

 & 10 & 1 & 0.68 & 1.15 \\

 & 1 & 1 & 0.78 & 1.27  \\

\tableline

B & 300 & 9 & 0.90 & 1.40-1.76 \\

 & 50 & 1 & 0.82 & 1.38 \\

 & 10 & 1 & 0.84 & 1.37 \\

 & 1 & 1 & 1.1 & 1.25   \\

\tableline

R & 50 & 1 & 0.64 & 1.14 \\

 & 10 & 1 & 0.66 & 1.13 \\

\enddata

%% Text for table notes should follow after the \enddata but before
%% the \end{deluxetable}. Make sure there is at least one \tablenotemark
%% in the table for each \tablenotetext.

%% If you use the table environment, please indicate horizontal rules using
%% Do not put multiple tabular environments within a single table.
%% The optional \label should appear inside the \caption command.

\end{deluxetable}

\clearpage

\begin{deluxetable}{cccc}
\tabletypesize{\scriptsize} \tablecaption{Annulus Geometry
\label{table2}} \tablewidth{0pt} \tablehead{\colhead{Annulus} &
\colhead{Radius ($'$)} & \colhead{Radius (pixels)} & \colhead{Area
($'$)$^{2}$}} \startdata

A1 & 0 $\leq$ R $\leq$ 0.5 & 0 $\leq$ R $\leq$ 145 & 0.51  \\

A2 & 0 $\leq$ R $\leq$ 1.0 & 0 $\leq$ R $\leq$ 291 & 2.59  \\

A3 & 0.5 $\leq$ R $\leq$ 1.5 & 145 $\leq$ R $\leq$ 437 & 5.74 \\

A1+A3 & 0 $\leq$ R $\leq$ 1.5 & 0 $\leq$ R $\leq$ 437 & 6.25 \\

A4 & 1.5 $\leq$ R $\leq$ 2.5 & 437 $\leq$ R $\leq$ 728 & 12.02 \\

A5 & 2.5 $\leq$ R $\leq$ 3.5 & 728 $\leq$ R $\leq$ 1019 & 18.30 \\

A6 & 3.5 $\leq$ R $\leq$ 4.5 & 1019 $\leq$ R $\leq$ 1311 & 24.58 \\

A7 & 4.5 $\leq$ R $\leq$ 5.5 & 1311 $\leq$ R $\leq$ 1602 & 30.87 \\

A8 & 5.5 $\leq$ R $\leq$ 6.5 & 1602 $\leq$ R $\leq$ 1893 & 37.15 \\

A9 & 6.5 $\leq$ R $\leq$ 7.5 & 1893 $\leq$ R $\leq$ 2184 & 43.43 \\

A10 & 7.5 $\leq$ R $\leq$ 8.5 & 2184 $\leq$ R $\leq$ 2475 & 49.72 \\

A11 & 8.5 $\leq$ R $\leq$ 9.5 & 2475 $\leq$ R $\leq$ 2767 & 55.98 \\

Global & 0 $\leq$ R $\leq$ 9.5 & 0 $\leq$ R $\leq$ 2767 & 278.31 \\

\enddata

%% Text for table notes should follow after the \enddata but before
%% the \end{deluxetable}. Make sure there is at least one \tablenotemark
%% in the table for each \tablenotetext.

%% If you use the table environment, please indicate horizontal rules using
%% Do not put multiple tabular environments within a single table.
%% The optional \label should appear inside the \caption command.

\end{deluxetable}

\clearpage

\begin{deluxetable}{ccccccc}
\tabletypesize{\scriptsize} \tablecaption{Completeness Corrections
\label{table3}} \tablewidth{0pt} \tablehead{\colhead{V mag} &
\colhead{No. Stars Input} & \colhead{No. Stars Recovered
(Cluster/Blank)} & \colhead{Completeness Correction
(Cluster/Blank)}} \startdata

Main-Sequence \\

14.5-15.0 &127 &122/123 &1.043/1.038  \\

15.0-15.5 &126 &122/124 &1.035/1.016  \\

15.5-16.0 &98 &93/96 &1.06/1.021  \\

16.0-16.5 &107 &103/107 &1.047/1.000  \\

16.5-17.0 &93 &84/88 &1.107/1.057  \\

17.0-17.5 &90 &87/88 &1.038/1.023  \\

17.5-18.0 &103 &92/96 &1.127/1.073  \\

18.0-18.5 &96 &85/91 &1.133/1.055  \\

18.5-19.0 &82 &76/80 &1.084/1.023  \\

19.0-19.5 &84 &72/76 &1.176/1.105  \\

19.5-20.0 &82 &70/75 &1.183/1.093  \\

20.0-20.5 &71 &56/60 &1.283/1.183  \\

20.5-21.0 &76 &58/63 &1.306/1.206  \\

21.0-21.5 &62 &49/50 &1.28/1.248  \\

21.5-22.0 &59 &42/44 &1.415/1.341  \\

22.0-22.5 &44 &27/31 &1.664/1.419  \\

22.5-23.0 &32 &19/21 &1.684/1.584  \\

23.0-23.5 &20 &11/12 &1.818/1.667  \\

\tableline

White Dwarfs \\

21.0-22.0 &28 &26.5/27 &1.057/1.037  \\

22.0-23.0 &46 &29.3/31 &1.570/1.484  \\

23.0-24.0 &51 &30.5/32 &1.672/1.594  \\

24.0-25.0 &59 &26/29 &2.269/2.034  \\

\enddata

%% Text for table notes should follow after the \enddata but before
%% the \end{deluxetable}. Make sure there is at least one \tablenotemark
%% in the table for each \tablenotetext.

%% If you use the table environment, please indicate horizontal rules using
%% Do not put multiple tabular environments within a single table.
%% The optional \label should appear inside the \caption command.

\end{deluxetable}

\clearpage

\begin{deluxetable}{ccccccccccc} \rotate
\tabletypesize{\scriptsize} \tablecaption{Cluster Star Counts (Raw
/ Corrected) \label{table4}} \tablewidth{0pt}
\tablehead{\colhead{V mag} & \colhead{A1+A3} & \colhead{A4} &
\colhead{A5} & \colhead{A6} & \colhead{A7} & \colhead{A8} &
\colhead{A9} & \colhead{A10} & \colhead{V mag} & \colhead{GLOBAL}}
\startdata

15.5-16.5 (Raw)  &63 & 91 &104 & 68 &49 &23  &15 &29& 15.0-16.0 &
441 \\

Corrected  &64.3 (8.4) &96.8 (11.3) & 110.6 (12.1) &72.6 (10.6) &
52.5 (9.4) &25.6 (9.8) & 17.1 (9.3) & 32.1 (10.6) && 459.5 (30.4)
\\
\\

16.5-17.5  &33 &51&56 &52&28 &21&39&15& 16.0-17.0 & 351 \\

Corrected  &35.4 (7.7) &57.1 (9.9) &62.7 (10.5) &58.5 (10.8)&33.2
(12.0) &25.9 (12.6)&46.4 (14.7)&20.2 (14.2)&&378.5 (32.5)
\\
\\

17.5-18.5 &20 &33&37 &61&40 &36&19&13& 17.0-18.0 & 315 \\

Corrected &22.4 (6.7) &38.0 (9.1)&43.2 (10.9) &71.3 (14.6)&47.7
(13.5) &44.4 (15.5)&25.6 (15.3)&19.5 (16.0)&&333.9 (40.5)
\\
\\

18.5-19.5 &14 &23&58 &35&53 &37&34&18& 18.0-19.0 & 315 \\

Corrected &15.1 (4.9) &25.5 (7.1)&63.7 (10.2) &39.2 (9.6)&59.5
(12.0) &42.5 (11.7)&39.5 (12.1)&22.8 (12.2)&&385.6 (46.3)
\\
\\

19.5-20.5 &3 &28&19 &19&36 &12&16&25& 19.0-20.0 & 209 \\

Corrected &3.2 (4.7) &33.5 (7.3)&24.4 (9.3) &25.0 (10.4) &46.3
(13.3) &19.2 (13.4)&25.1 (15.1)&36.4 (16.5)&&263.1 (40.3)
\\
\\

20.5-21.5 &6 &23&20 &22&43 &14&18&34& 20.0-21.0 & 201 \\

Corrected &8.6 (4.7) &31.4 (9.7)&29.1 (12.2) &32.1 (13.1)&61.3
(17.5) &25.1 (17.4)&30.9 (18.6)&52.7 (21.0)&&293.4 (59.7)
\\
\\

21.5-22.5 &0 &18&8 &35&18 &29&2&13& 21.0-22.0 & 193 \\

Corrected &0 &26.8 (11.2)&14.4 (10.1) &53.1 (19.3)&32.5 (25.9)
&49.9 (31.3)&13.0 (10.3)&29.8 (15.4)&&252.8 (87.9)
\\
\\

22.5-23.5 &0 &0&3 &24&15 &17&25&0& 22.0-23.0 & 169 \\

Corrected &0 &0&10.0 (7.0) &47.1 (20.1)&37.5 (17.8) &42.0 (18.1)
&56.7 (14.9)&0 (6.8)&&286.6 (181.8)
\\

\enddata

%% Text for table notes should follow after the \enddata but before
%% the \end{deluxetable}. Make sure there is at least one \tablenotemark
%% in the table for each \tablenotetext.

%% If you use the table environment, please indicate horizontal rules using
%% Do not put multiple tabular environments within a single table.
%% The optional \label should appear inside the \caption command.

\end{deluxetable}

\clearpage

\begin{deluxetable}{ccccccc}
\tabletypesize{\scriptsize} \tablecaption{Equal Mass Binary Star
Counts \label{table5}} \tablewidth{0pt} \tablehead{\colhead{Ann
No.} & \colhead{Radius ($'$)} & \colhead{No. MS Stars (15.5 $\leq$
V $\leq$ 21.5)} & \colhead{No. Binary Stars (15.5 $\leq$ V $\leq$
21.5)} & \colhead{Percentage of Binaries (\%)}} \startdata

A1 &0 $\leq$ R $\leq$ 0.5 &12 ($\pm$6) &2 ($\pm$1) & 17 ($\pm$12) \\

A2 &0 $\leq$ R $\leq$ 1 &71 (9) &3 (4) & 4 (6) \\

A3 &0.5 $\leq$ R $\leq$ 1.5 &191 (14)&12 (6) & 6 (3)\\

A1+A3 &0 $\leq$ R $\leq$ 1.5 &203 (15) &13 (6) &6 (3)\\

A4 &1.5 $\leq$ R $\leq$ 2.5 &291 (20) &34 (8) &12 (3)\\

A5 &2.5 $\leq$ R $\leq$ 3.5 &319 (21) &48 (10) &15 (3)\\

A6 &3.5 $\leq$ R $\leq$ 4.5 &287 (21) &36 (11) &13 (4)\\

A7 &4.5 $\leq$ R $\leq$ 5.5 &275 (22) &57 (13) &21 (5)\\

A8 &5.5 $\leq$ R $\leq$ 6.5 &160 (26) &6 (13)  &4 (8)\\

A9 &6.5 $\leq$ R $\leq$ 7.5 &144 (26) &26 (15)  &18 (11) \\

A10 &7.5 $\leq$ R $\leq$ 8.5 &110 (29) &28 (16) &25 (16) \\

Global &0 $\leq$ R $\leq$ 9.5 &2214 (67) &242 (37.5) &11 (2) \\

\enddata

%% Text for table notes should follow after the \enddata but before
%% the \end{deluxetable}. Make sure there is at least one \tablenotemark
%% in the table for each \tablenotetext.

%% If you use the table environment, please indicate horizontal rules using
%% Do not put multiple tabular environments within a single table.
%% The optional \label should appear inside the \caption command.

\end{deluxetable}

\clearpage

\begin{deluxetable}{cccccc}
\tabletypesize{\scriptsize} \tablecaption{White Dwarf Continuity
Analysis -- Predicted Number vs. Observed Number \label{table6}}
\tablewidth{0pt} \tablehead{\colhead{V mag cut} &
\colhead{Observed (Raw)} & \colhead{Observed (Corr)} &
\colhead{Pred (0.6 M$_\odot$)} & \colhead{Pred (0.7 M$_\odot$)} &
\colhead{Pred (0.8 M$_\odot$)}} \startdata

For 0.80 Stellarity Cut \\ 13.0$\pm$3.6 Red Giants \\ $t_{RG}$ = 9
$\times$ 10$^{7}$
\\

$\leq$24   & 27 ($\pm$5) & 53 ($\pm$7) & 61 ($\pm$17) & 55
($\pm$16) & 48 ($\pm$14) \\

$\leq$23.5 & 21 (5) & 35 (6) & 29 (8) & 25 (7) & 21 (6) \\

$\leq$23   & 17 (4) & 28 (6) & 11 (3) & 9 (3) & 8 (2) \\
\tableline

For $t_{RG}$ = 7 $\times$ 10$^{7}$
\\

$\leq$24   & 27 (5) & 53 (7) & 78 (22) & 71 (20) & 61 (17) \\

$\leq$23.5 & 21 (5) & 35 (6) & 37 (10) & 32 (9) & 27 (8) \\

$\leq$23   & 17 (4) & 28 (6) & 15 (4) & 12 (3) & 10 (3) \\
\tableline

For $t_{RG}$ = 5 $\times$ 10$^{7}$
\\

$\leq$24   & 27 (5) & 53 (7) & 109 (31) & 99 (28) & 86 (24) \\

$\leq$23.5 & 21 (5) & 35 (6) & 52 (15) & 44 (13) & 38 (11) \\

$\leq$23   & 17 (4) & 28 (6) & 20 (6) & 17 (5) & 14 (4) \\
\tableline \tableline

For 0.90 Stellarity Cut \\ $t_{RG}$ = 9 $\times$ 10$^{7}$
\\

$\leq$24   & 17 (4) & 34 (6) & 61 (17) & 55 (16) & 48 (14) \\

$\leq$23.5 & 13 (4) & 22 (5) & 29 (8) & 25 (7) & 21 (6) \\

$\leq$23   & 12 (3) & 19 (5) & 11 (3) & 9 (3) & 8 (2) \\
\tableline

For $t_{RG}$ = 7 $\times$ 10$^{7}$
\\

$\leq$24   & 17 (4) & 34 (6) & 78 (22) & 71 (20) & 61 (17) \\

$\leq$23.5 & 13 (4) & 22 (5) & 37 (10) & 32 (9) & 27 (8) \\

$\leq$23   & 12 (3) & 19 (5) & 15 (4) & 12 (3) & 10 (3) \\
\tableline

For $t_{RG}$ = 5 $\times$ 10$^{7}$
\\

$\leq$24   & 17 (4) & 34 (6) & 109 (31) & 99 (28) & 86 (24) \\

$\leq$23.5 & 13 (4) & 22 (5) & 52 (15)  & 44 (13) & 38 (11) \\

$\leq$23   & 12 (3) & 19 (5) & 20 (6)  & 17 (5)  & 14 (4) \\

\enddata

%% Text for table notes should follow after the \enddata but before
%% the \end{deluxetable}. Make sure there is at least one \tablenotemark
%% in the table for each \tablenotetext.

%% If you use the table environment, please indicate horizontal rules using
%% Do not put multiple tabular environments within a single table.
%% The optional \label should appear inside the \caption command.

\end{deluxetable}

\clearpage

%% You can append references to a table using the \tablerefs command.

%% Tables may also be prepared as separate files. See the accompanying
%% sample file table.tex for an example of an external table file.
%% To include an external file in your main document, use the \input
%% command. Uncomment the line below to include table.tex in this
%% sample file.

%\input{table}

%% The following command ends your manuscript. LaTeX will ignore any text

\end{document}